\documentclass[reprint, notitlepage, superscriptaddress, amsmath, amssymb, aps, prl]{revtex4-2}
\usepackage{color}
\usepackage{graphicx}
\usepackage{dcolumn}
\usepackage{bm}
\usepackage{hyperref}
\usepackage{braket}
\usepackage{mathtools}
\usepackage{enumerate}
\usepackage{MnSymbol}

\newcommand{\nspace}[1]{\hspace*{#1em}} 

\begin{document}

\title{Superconducting nonlinear Hall effect induced by geometric phases}

\author{Kazuaki Takasan}
\email{kazuaki.takasan@phys.s.u-tokyo.ac.jp}
\affiliation{Department of Physics, University of Tokyo, 7-3-1 Hongo, Tokyo 113-0033, Japan}

\author{Naoto Tsuji}
\email{tsuji@phys.s.u-tokyo.ac.jp}
\affiliation{Department of Physics, University of Tokyo, 7-3-1 Hongo, Tokyo 113-0033, Japan}
\affiliation{RIKEN Center for Emergent Matter Science (CEMS), 2-1 Hirosawa, Wako, Saitama 351-0198, Japan}

\date{\today}

\begin{abstract}
    We study the nonlinear Hall effect in superconductors without magnetic fields induced by a quantum geometric phase (i.e., the Aharonov-Bohm phase) carried by single or pair particles. We find that the second-order nonlinear Hall conductivity diverges in the dc limit in a robust way against dissipation when the system is superconducting, suggesting that the supercurrent flows perpendicular to the direction of the applied electric field. This superconducting nonlinear Hall effect (SNHE) is demonstrated for the Haldane model with attractive interaction and its variant with pair hoppings. 
    In the Ginzburg-Landau theory, the SNHE can be understood as those arising from a higher-order Lifshitz invariant, that is, a symmetry invariant constructed from order parameters that contains an odd number of spatial derivatives. We perform real-time simulations including the effect of collective modes for the models driven by a multi-cycle pulse, and show that the SNHE leads to large rectification of the Hall current under light driving.
\end{abstract}

\maketitle

\textit{Introduction.}---
Electromagnetic responses have long been a central subject in the research of superconductors. 
Despite its long history of investigations~\cite{Tinkham_book}, intriguing new phenomena continue to be discovered, including 
nonlinear optical responses mediated by the Higgs mode in superconductors~\cite{Matsunaga2014, Katsumi2018, Chu2020, Higgs_review} and nonreciprocal critical currents that give rise to the superconducting diode effect~\cite{Ando2020, Baumgartner2022, Nadeem2023}. Investigating novel electromagnetic responses is crucial not only 
for studying microscopic mechanisms of superconductivity but also for exploring future device technologies for electronics and spintronics~\cite{Braginski2019,Linder2015}.

Among a wide range of electromagnetic responses, the Hall effect, in which a current flows perpendicular to the applied voltage, has been extensively studied in both fundamental and applied contexts~\cite{Hall1879,Nagaosa2010, Sinova2015, Klitzing2020}. 
Studies of the Hall effect have been broadened to various related phenomena, such as the anomalous Hall effect~\cite{Nagaosa2010}, spin Hall effect~\cite{Sinova2015}, and quantum Hall effect~\cite{Klitzing2020}. More recently, the exploration has been expanded to the nonlinear regime, whose studies on fundamental aspects and possible applications are still rapidly developing~\cite{Ma2019, Kang2019, Du2021}.

\begin{figure}
    \centering
    \includegraphics[width=8cm]{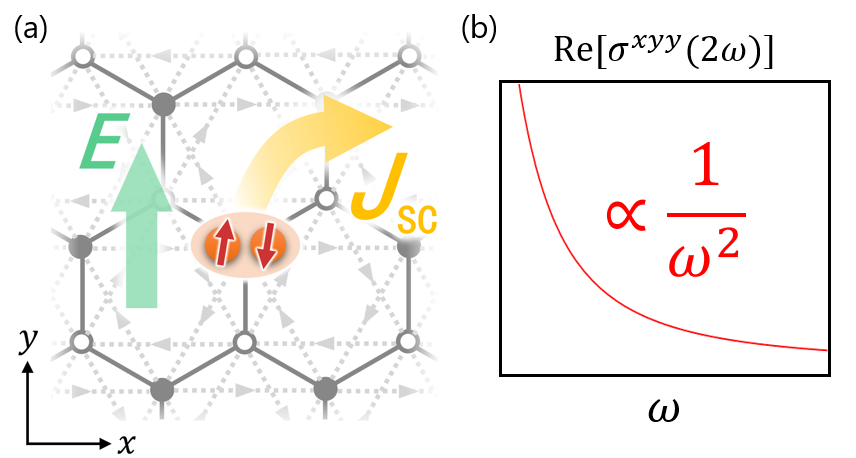}
    \caption{(a) Schematic illustration of the superconducting nonlinear Hall effect driven by geometric phases. The black (white) circles represent sites in the sublattice $\mathbb{A}$ ($\mathbb{B}$), the dashed lines with arrows represent single or pair hoppings attached with the geometric phase, and $E$ and $J_{\rm sc}$ denote the applied electric field and the induced supercurrent flow, respectively. (b) Divergent behavior of the nonlinear Hall conductivity in the dc limit, which characterizes the superconducting nonlinear Hall effect.}
    \label{fig:concenpt}
\end{figure}

For superconductors, most of the previous studies on the Hall effect have focused on the responses in the presence of magnetic fields arising from vortex motion, where normal carriers inside vortex cores lead to a dissipative current~\cite{Tinkham_book, Bardeen1965, Caroli1967, Blatter1994}. This raises the question of whether it is possible to realize a dissipationless Hall current (i.e., transverse supercurrent flow) in superconductors (see Fig.~\ref{fig:concenpt}(a)).
While the corresponding divergent Hall conductivity does not appear in the linear response regime near equilibrium~\footnote{The divergent contribution of the linear optical conductivity $\sigma^{\mu\nu}(\omega)$ is proportional to $ (\partial_{A_\mu} \partial_{A_\nu} \mathcal{F})|_{A=0}$, where $\mathcal{F}$ is the free energy and $A_\mu$ is the vector potential. Since the derivatives are commutative, $\sigma^{xy}(\omega)=\sigma^{yx}(\omega)$ for the divergent component and there is no Hall contribution.}, recent theoretical studies have revealed that, in the nonlinear regime, the Hall conductivity can diverge in the low-frequency limit~\cite{Watanabe2022, Tanaka2023, Tanaka2024, Matsyshyn2024arXiv}. We refer to this phenomenon as the superconducting nonlinear Hall effect (SNHE) (Fig.~\ref{fig:concenpt}(b)). Despite its nonlinear nature, one could observe a gigantic Hall response if the nonlinear Hall conductivity would diverge. We mainly focus on the maximal divergence proportional to $1/\omega^2$, which is only possible if the time-reversal symmetry is broken, as shown later.

Compared to the case of normal states,
the microscopic mechanism of the SNHE is largely unexplored. In particular, 
some of the previous studies have focused on unconventional pairing symmetries (e.g., $s+p$ or $s+ip$ pairing)~\cite{Watanabe2022, Tanaka2023}, superconductors in magnetic fields \cite{Tanaka2024}, and time-reversal symmetric superconductors \cite{Matsyshyn2024arXiv}. So far, dissipation has been modeled only phenomenologically,
whereas its microscopic treatment is crucial to rule out a spurious divergence in clean metallic systems without superconducting orders. 
Various other types of (non-divergent) nonlinear Hall effects in superconductors have been recently investigated theoretically~\cite{Parafilo2023, Daido2024, Sonowal2024, Mironov2024, Dong2024arXiv} and experimentally \cite{Itahashi2022}.

In this paper, we propose an intriguing mechanism for the SNHE in time-reversal symmetry broken $s$-wave superconductors without magnetic fields. The key ingredient is a quantum geometric phase (i.e., the Aharanov-Bohm (AB) phase) carried by single (or pair) particles.
The geometric phase has played a pivotal role in transport, dynamics, and topology of quantum systems \cite{Bohm_book, Xiao2010, Cohen2019}, allowing one to effectively insert microscopic fluxes with keeping zero net flux \cite{Haldane1988}. This is advantageous in the current context, since a net magnetic flux is expelled in superconductors while microscopic AB phases can be induced in various ways, which will be discussed later.

To demonstrate the mechanism, we study a honeycomb lattice model that incorporates two kinds of AB phases: one attached to single-particle hoppings, similar to that of the Haldane model for Chern insulators \cite{Haldane1988}, and another associated with two-particle hoppings. 
By calculating with the Keldysh Green's functions, which include the effect of dissipation microscopically, we reveal a robust divergence in the dc limit of the nonlinear Hall conductivity. We also discuss the SNHE from the Ginzburg–Landau framework, and point out that SNHEs are classified with higher-order Lifshitz invariants \cite{Landau_StatPhys_textbook}, which are symmetry invariants containing odd-order spatial derivatives. Furthermore, we employ a real-time simulation under light driving to show that the SNHE leads to the divergently strong rectification in the Hall current. 

\textit{Model.}--- 
We consider a two-dimensional $s$-wave superconductor on the honeycomb lattice ($2L^2$ sites), described by the Hamiltonian,
\begin{align}
    \hat{H}_0 &=\! - t_1 \!\!\!\! \sum_{\langle i,j \rangle, \sigma} \!\!\!\! \left( \hat{c}^\dagger_{i,\sigma} \hat{c}_{j,\sigma} + \mathrm{h.c.} \right) - U \sum_{i} \hat{n}_{i,\uparrow} \hat{n}_{i,\downarrow} \nonumber \\
    &\quad + (-\mu + m) \!\! \sum_{i \in \mathbb{A}, \sigma}\!\!  \hat{n}_{i,\sigma} + (-\mu - m)\!\! \sum_{i \in \mathbb{B}, \sigma}\!\!  \hat{n}_{i,\sigma}, \label{eq:H0}  
\end{align}
where $\hat{c}_{i,\sigma}$ and $\hat{n}_{i,\sigma} = \hat{c}^\dagger_{i,\sigma} \hat{c}_{i,\sigma}$ are the annihilation and number operators for electrons with spin $\sigma$ at site $i$, respectively, and $\langle i,j \rangle$ denotes a pair of nearest-neighbor sites. We set $t_1=1$ as the unit of energy throughout this study. On top of the chemical potential $\mu$, we introduce a staggered potential $m$  breaking inversion symmetry, where the lattice consists of two sublattices $\mathbb{A}$ and $\mathbb{B}$ as shown in Fig.~\ref{fig:concenpt}(a). To study $s$-wave superconductors, we consider an attractive Hubbard interaction $-U<0$, and decouple it with the mean-field order parameter $\Delta_i = - U \langle \hat{c}_{i, \downarrow} \hat{c}_{i, \uparrow} \rangle$. For simplicity, we assume that the order parameter is uniform in each sublattice, i.e., $\Delta_i = \Delta_\mathrm{A} (\Delta_\mathrm{B})$ for $i \in \mathbb{A} (\in \mathbb{B})$. 

To investigate the effect of geometric phases, we introduce an AB phase $\phi$ in two different manners. First one is the single-particle geometric phase, which is attached to the single-particle hopping term in the Hamiltonian,
\begin{equation}
    \hat{H}_1 = - t_2 \!\!\!\! \sum_{\llangle i,j \rrangle, \sigma} \!\!\!\! \left(e^{i\phi}  \hat{c}^\dagger_{i,\sigma} \hat{c}_{j,\sigma} + \mathrm{h.c.}\right), \label{eq:H1}
\end{equation}
where $\llangle i,j \rrangle$ denotes a pair of next nearest-neighbor sites,
and the sign of $\phi$ is taken to be positive in the direction of the arrows in Fig.~\ref{fig:concenpt}(a).
This is the same as the one in the Haldane model showing the quantum anomalous Hall effect~\cite{Haldane1988}. The second one is the two-particle AB phase, which is attached to the pair hopping terms,
\begin{align}
    \hat{H}_2 &= - J_1 \sum_{\langle i,j \rangle}  \left( \hat{c}^\dagger_{i,\uparrow} \hat{c}^\dagger_{i,\downarrow} \hat{c}_{j,\downarrow} \hat{c}_{j,\uparrow} + \mathrm{h.c.} \right) \nonumber \\
    & \quad - J_2 \sum_{\llangle i,j \rrangle} \left( e^{2i\phi} \hat{c}^\dagger_{i,\uparrow} \hat{c}^\dagger_{i,\downarrow} \hat{c}_{j,\downarrow} \hat{c}_{j,\uparrow} + \mathrm{h.c.} \right). \label{eq:H2}
\end{align}
Note that the AB phase is doubled reflecting the charge of electron pairs. This type of the pair hopping has been considered in the Penson-Kolb model~\cite{Penson1986, Robaszkiewicz1999, Czart2001}. We adopt the mean-field approximation, 
$\hat{c}^\dagger_{i,\uparrow} \hat{c}^\dagger_{i,\downarrow} \hat{c}_{j,\downarrow} \hat{c}_{j,\uparrow} 
\to \Delta_i^* \hat{c}_{i,\uparrow} \hat{c}_{i,\downarrow} + \Delta_i \hat{c}^\dagger_{j,\downarrow} \hat{c}^\dagger_{j,\uparrow} - \Delta_i^* \Delta_j$, with which we use $\hat{H}_\mathrm{sys} = \hat{H}_0 + \hat{H}_1 + \hat{H}_2$ 
as the system Hamiltonian. 

To take into account the effect of dissipation in the calculation of the nonlinear conductivity, we consider the coupling to a heat bath at temperature $T$, described by a Hamiltonian $\hat{H}_\mathrm{tot} = \hat{H}_\mathrm{sys} + \hat{H}_\mathrm{bath} + \hat{H}_\mathrm{int}$ where $\hat{H}_\mathrm{bath} = \sum_{i, n, \sigma} \varepsilon_n b_{i,n,\sigma}^\dagger b_{i,n,\sigma}, \hat{H}_\mathrm{int} = \sum_{i,n,\sigma} V_{n} \hat{c}_{i,\sigma}^\dagger \hat{b}_{i,n,\sigma} + \mathrm{h.c.}$ \cite{Tsuji2009}. Here, $b_{i,n,\sigma}$ is the fermionic annihilation operator for the $n$-th mode of the heat bath at the $i$-th site, and $V_{n}$ is the coupling strength. The effect of the bath is taken into account with the self-energy in the Keldysh formalism, which results in a finite lifetime $\propto 1/\Gamma$ for electrons~\footnote{For detail of the theoretical treatment of the heat bath, see Supplemental Material \ref{sec:supp_bath}.}.

For numerical calculation, we mainly consider two parameter sets: (i) $(t_2, J_1, J_2)=(0.2, 0, 0)$ and (ii) $(t_2, J_1, J_2)=(0, 0.5, 0.2)$. We refer to a superconducting phase with the parameter (i) [(ii)] with $U=4$ as \textsf{SC1} [\textsf{SC2}]. For normal phases, we consider the cases of (i) and (ii) with setting $\Delta_\mathrm{A} = \Delta_\mathrm{B}=0$, and they are referred to as \textsf{NM1} and \textsf{NM2}, respectively. For all the cases, we set $\mu=-0.8$ and $m=0.5$. 

\textit{Nonlinear conductivity.}---
To investigate the electromagnetic responses, we consider an external field $\bm A(t)$ introduced via the Peierls phase, $\hat{c}_{i,\sigma} \rightarrow \hat{c}_{i,\sigma} e^{i \bm A(t) \cdot \bm R_i}$, where $\bm R_i$ is the position vector of site $i$, to Eqs.~\eqref{eq:H1}-\eqref{eq:H2}. 
The second-order conductivity $\sigma^{\mu \nu \lambda}(\omega_{12}; \omega_{1}, \omega_{2})$~($\mu, \nu, \lambda=x, y$) is defined via $j^{\mu(2)}(\omega) = \sum_{\nu \lambda}  \int \frac{d\omega_1}{2\pi} \int \frac{d\omega_2}{2\pi} \sigma^{\mu \nu \lambda}(\omega_{12}; \omega_1, \omega_2)  2\pi \delta(\omega - \omega_{12}) 
E_\nu(\omega_1)E_\lambda(\omega_2)$, where $\omega_{12}=\omega_1+\omega_2$, and $j^{\mu(2)}(\omega)$ and $E_\mu(\omega)$ are the Fourier components of the second-order current density and the external electric field $\bm E(t) [= - \partial_t \bm A(t)]$, respectively. Based on the perturbative expansion with the Keldysh Green's function~\cite{Kamenev_book} including the effect of dissipation to the heat bath~\footnote{The outline of derivation is presented in Supplemental Material \ref{sec:supp_derivation_nonlinear_cond}.}, we obtain the general expression for the nonlinear conductivity as follows:
\begin{align}
        &\sigma^{\mu \nu \lambda}(\omega_{12}; \omega_{1}, \omega_2) \nonumber \\
        &~\!=
                \frac{1}{2\omega_1 \omega_2}\! \frac{1}{L^2}\sum_{\bm k }\!
                 \int\!\frac{d\Omega}{2\pi i} f(\Omega) \mathrm{Tr} \Big[  \Pi^{\mu \nu \lambda}_{1\mathrm{ph}}(\Omega) + \Pi^{\mu \nu \lambda}_{2\mathrm{ph}}(\Omega) \nonumber \\
                 &\quad+\! \Pi^{\mu \nu \lambda}_{3\mathrm{ph}}(\Omega)  +\! [(\omega_1, \nu)\! \leftrightarrow \! (\omega_2, \lambda)] \Big]~\!\!\!+\! \frac{1}{2\omega_1 \omega_2}\frac{1}{L^2} \mathcal{E}_0^{\mu\nu\lambda}, \label{eq:formula_conductivity}
\end{align}
\vspace{-0.5cm}
\begin{align*}
    \Pi^{\mu \nu \lambda}_{1\mathrm{ph}}(\Omega) &= V^{\mu}_{\bm k}  G_{\bm k }^{R}(\Omega + \omega_{12}) V_{\bm k}^\nu G_{\bm k }^{R}(\Omega + \omega_2) V_{\bm k}^\lambda \delta G_{\bm k }(\Omega) \\ 
    &\quad +  V^{\mu}_{\bm k} G_{\bm k}^{R}(\Omega + \omega_{1}) V_{\bm k}^{\nu} \delta G_{\bm k }(\Omega) V_{\bm k}^{\lambda} G_{\bm k }^{A}(\Omega - \omega_2) \\
    &\quad +V^{\mu}_{\bm k} \delta G_{\bm k }(\Omega) V_{\bm k}^\nu G_{\bm k}^{A}(\Omega-\omega_1) V_{\bm k}^\lambda  G_{\bm k }^{A}(\Omega- \omega_{12}), \\
    \Pi^{\mu \nu \lambda}_{2\mathrm{ph}}(\Omega) &= \frac{1}{2} \Big\{  
    V^{\mu}_{\bm k} G_{\bm k }^{R}(\Omega + \omega_{12}) W_{\bm k}^{\nu \lambda}  \delta G_{\bm k }(\Omega)  - G^{R}_{\bm k }(\Omega) \\
    &\qquad + V^{\mu}_{\bm k}  \delta G_{\bm k }(\Omega) W_{\bm k}^{\nu \lambda} G_{\bm k }^{A}(\Omega - \omega_{12}) \Big \} \\
    &\quad + W_{\bm k}^{\mu \nu} G^{R}_{\bm k }(\Omega+\omega_2)  V_{\bm k}^\lambda  \delta G_{\bm k }(\Omega) \\
    &\qquad + W_{\bm k}^{\mu \nu} \delta G_{\bm k }(\Omega) V_{\bm k}^\lambda G^{A}_{\bm k }(\Omega - \omega_2),
\end{align*}
$\Pi^{\mu \nu \lambda}_{3\mathrm{ph}}(\Omega)= \frac{1}{2}X^{\mu \nu \lambda}_{\bm k} \delta G_{\bm k }(\Omega)$, where  $f(\Omega)=1/(e^{\Omega/T}+1)$, $G_{\bm k}^{R/A}(\Omega)=(\Omega - h^\mathrm{BdG}_{\bm k, \bm 0} \pm i \Gamma)^{-1}$\footnote{Note that $\Gamma$, which represents the coupling strength to heat bath, does not have to be infinitesimally small, while it is similar to the infinitesimal imaginary part in linear response theory.}, $\delta G_{\bm k}(\Omega) = G_{\bm k}^{A}(\Omega) - G_{\bm k}^{R}(\Omega)$, $V_{\bm k}^{\mu}= \frac{\partial h^\mathrm{BdG}_{\bm k, \bm A}}{\partial A_{\mu}}|_{A=0}$, $W_{\bm k}^{\mu \nu}=\frac{\partial^2 h^\mathrm{BdG}_{\bm k, \bm A}}{\partial A_{\mu} \partial A_{\nu}}|_{A=0}$, and $X^{\mu \nu \lambda}_{\bm k}= \frac{\partial^3 h^\mathrm{BdG}_{\bm k, \bm A}}{\partial A_{\mu} \partial A_{\nu} \partial A_{\lambda}}|_{A=0}$ with the Bogoliubov-de Gennes Hamiltonian
$h^\mathrm{BdG}_{\bm k, \bm A} =
{\scriptsize \begin{pmatrix}
    h^\mathrm{N}_{\bm k - \bm A} & \Delta_{\bm A} \\
    \Delta^\dagger_{\bm A} & -(h^\mathrm{N}_{-\bm k - \bm A})^\mathsf{T}
\end{pmatrix}}$,
which consists of the normal part
$h^\mathrm{N}_{\bm k} = {\scriptsize \begin{pmatrix}
        t_2 f_2(\bm k, \phi) & t_1 f_1(\bm k) \\
        t_1 f^*_1(\bm k) & t_2 f_2(\bm k, -\phi)
    \end{pmatrix}}$,
and the anomalous part $\Delta_{\bm A}\!=\!\mathrm{diag}[\Delta_1(\bm A), \Delta_2(\bm A)]$, and $\mathcal{E}_0^{\mu\nu\lambda} = \frac{\partial ^3 \mathcal{E}_{0, \bm A}}{\partial A_{\mu} \partial A_{\nu} \partial A_{\lambda}}|_{A=0}$ where $\mathcal{E}_{0, \bm A}$ is the energy shift coming with the mean-field approximation. For the explicit form of the functions $f_{1,2}, \Delta_{1,2}$ and $\mathcal{E}_{0, \bm A}$, see the footnote~\footnote{We summarize the explicit forms of various functions used in our paper: $f_1(\bm k) = -\sum_{j=1}^3 e^{i \bm k \cdot \bm \alpha_j}$, $f_2(\bm k, \phi) = -2 \sum_{j=1}^3 \cos(\bm k \cdot \bm \beta_j + \phi)$ [Here, $\bm \alpha_1 = (0,1/\sqrt{3})$, $\bm \alpha_2 = (-1/2,-1/(2\sqrt{3}))$, $\bm \alpha_3 = (1/2,-1/(2\sqrt{3}))$, $\bm \beta_1 = (1, 0)$, $\bm \beta_2 = (-1/2, -\sqrt{3}/2)$, $\bm \beta_3 = (-1/2, \sqrt{3}/2)$], $\Delta_1(\bm A) = [  1 - \frac{J_2}{U} f_2(-2 \bm A, 2\phi) ]\Delta_{\mathrm{A}} - \frac{J_1}{U} f_1(-2 \bm A) \Delta_{\mathrm{B}}$, $\Delta_2(\bm A)  = [ 1 - \frac{J_2}{U} f_2(-2 \bm A, -2\phi)] \Delta_{\mathrm{B}} - \frac{J_1}{U} f_1^*\!(-2 \bm A) \Delta_{\mathrm{A}}$ and $\mathcal{E}_{0, \bm A} = \frac{L^2}{U} [1 - \frac{J_2}{U} f_2(-2 \bm A, 2\phi)]|\Delta_\mathrm{A}|^2 + \frac{L^2}{U}[ 1 - \frac{J_2}{U} f_2(-2 \bm A, -2\phi)]|\Delta_\mathrm{B}|^2-\frac{2 J_1 L^2 }{U^2} \mathrm{Re}[f_1(-2\bm A)  \Delta^{*}_\mathrm{A}\Delta_\mathrm{B}]$.}. 
Note that the effect of collective modes, which is expected to be negligible in the low-frequency regime far from the resonance \footnote{In fact, all the collective modes generally have an energy gap in ordinary $s$-wave superconductors.}, is ignored in Eq.~\eqref{eq:formula_conductivity}. Indeed, we will see later that the important features are unchanged even when we consider collective modes.

\begin{figure}
    \centering
    \includegraphics[width=7.8cm]{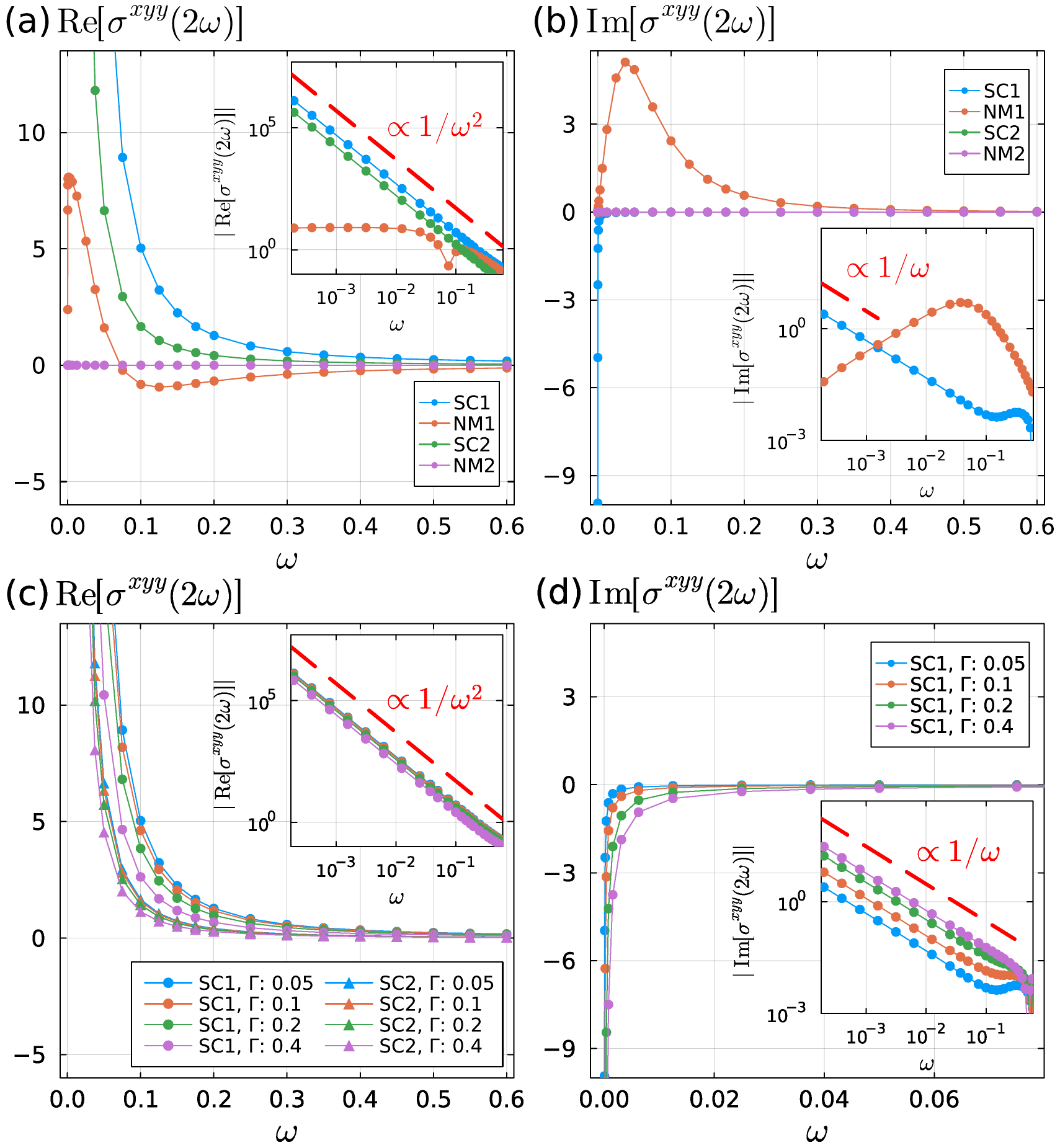}
    \caption{Low-frequency behavior of the nonlinear Hall conductivity $\sigma^{xyy}(2\omega; \omega, \omega)$ for the real [(a), (c)] and imaginary [(b), (d)] parts. (a) and (b) show the differences between the superconducting and normal phases. \textsf{SC1} (\textsf{SC2}) represents the superconducting phase with a single-particle (two-particle) geometric phase. \textsf{NM1} and \textsf{NM2} denote the corresponding normal phases. Here, we set $\Gamma = 0.05$.  (c) and (d) show the dependence on the dissipation strength $\Gamma$. Here, we omit \textsf{SC2}, in which the imaginary part is zero. We set $\phi=\pi/3$, $T=0.05$ and $L=120$ for all the plots. 
    Each inset is the log-log plot for the same data with an eye guide for $1/\omega$ or $1/\omega^2$.} 
    \label{fig:freq_dep}
\end{figure}

\textit{Divergence of the nonlinear Hall conductivity.}--- Figures \ref{fig:freq_dep}(a) and (b) show the low-frequency behavior of the nonlinear Hall conductivity $\sigma^{xyy}(2\omega; \omega, \omega)$ calculated with Eq.~\eqref{eq:formula_conductivity}. For both types of the geometric phases [Eq.~\eqref{eq:H1} and \eqref{eq:H2}], the Hall conductivity diverges in the dc limit in the superconducting phase ($\Delta_\mathrm{A}, \Delta_\mathrm{B} \neq 0$) [\textsf{SC1} and \textsf{SC2} in Fig.~\ref{fig:freq_dep}], while it remains finite in the normal phase ($\Delta_\mathrm{A}, \Delta_\mathrm{B} = 0$) [\textsf{NM1} and \textsf{NM2} in Fig.~\ref{fig:freq_dep}]. This is a clear signature of the SNHE,  
which occurs either by the single-particle or two-particle geometric phase. For the latter, we can analytically show in the clean limit that $\sigma^{xyy}(2\omega)\propto \frac{J_2}{U^2\omega^2}(|\Delta_A|^2-|\Delta_B|^2)\sin(2\phi)$~\footnote{The derivation is presented in Supplemental Material \ref{sec:supp_ana_derivation}.}. 
While the behavior of the real part is similar for both of the geometric phases as shown in Fig.~\ref{fig:freq_dep}(a), the imaginary part shows a sharp difference in Fig.~\ref{fig:freq_dep}(b): the divergence under the two-particle phase is absent, while it appears in the single-particle phase. This is because the two-particle phase does not induce the dc Hall conductivity in the linear response, which is directly related to this divergence in the imaginary part. For more details, see footnote~\footnote{The divergence in the imaginary part is related to the coefficient $\mathcal{B}^{xyy}$ introduced in Eq.~\eqref{eq:div_coeff} and it has a relationship with the linear DC Hall conductivity $\sigma^{xy}_{\mathrm{DC}, \bm A}$ under the vector potential $\bm A$ as $\mathcal{B}^{xyy} = i \partial_{A_y} \sigma^{xy}_{\mathrm{DC}, \bm A}|_{A=0}$~\cite{Watanabe2022}.}.

To confirm the robustness of the SNHE against dissipation, we plot $\sigma^{xyy}(2\omega; \omega, \omega)$ for various values of the dissipation strength $\Gamma$ in Fig.~\ref{fig:freq_dep}(c, d). We find that the divergence persists even with increasing $\Gamma$, which indicates robustness, while the value of the conductivity is modified. To discuss this change more precisely, we introduce the coefficients of the divergent terms, $\mathcal{A}^{xyy}$ and $\mathcal{B}^{xyy}$, as
\begin{equation}
    \sigma^{xyy}(2\omega; \omega, \omega) = \frac{\mathcal{A}^{xyy}}{\omega^2} + \frac{2\mathcal{B}^{xyy}}{\omega}  + \mathrm{reg.}, \label{eq:div_coeff}  
\end{equation}
where ``$\mathrm{reg.}$'' denotes the non-divergent part. The results of Fig.~\ref{fig:freq_dep}(c, d) indicate that $\mathcal{A}^{xyy} \in \mathbb{R}$ and $\mathcal{B}^{xyy} \in i \mathbb{R}$ remain finite with increasing $\Gamma$. In contrast to $\mathcal{A}^{xyy}$ suppressed with increasing $\Gamma$, $\mathcal{B}^{xyy}$ is enhanced. This suggests that the SNHE in the imaginary part is more robust than in the real part. 

\begin{figure}[t]
    \includegraphics[width=8.4cm]{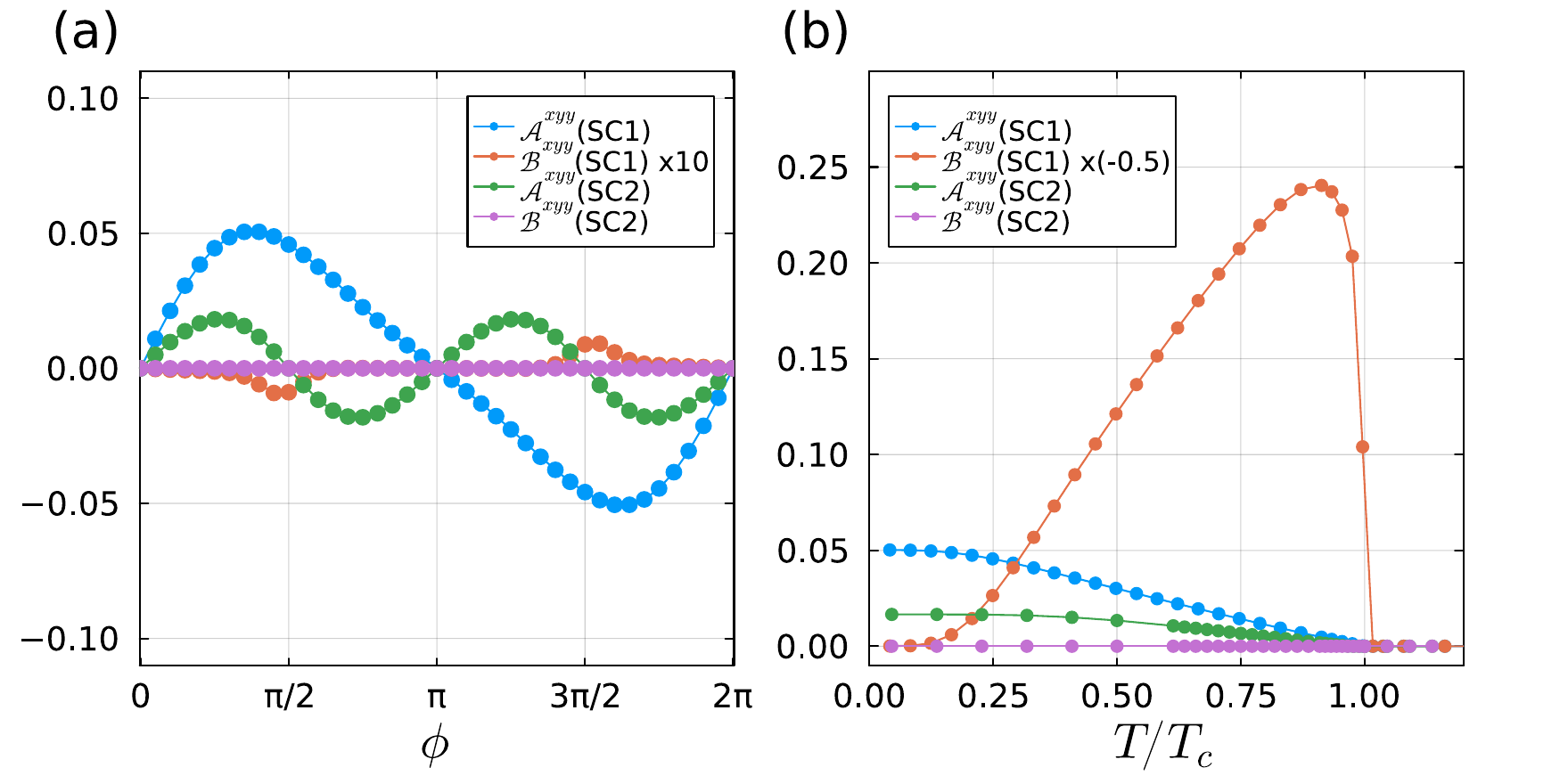}
    \caption{(a) $\phi$-dependence and (b) $T$-dependence of the coefficients $\mathcal{A}^{xyy}$ and $\mathcal{B}^{xyy}$ of the divergent terms (proportional to $1/\omega^2$ and $1/\omega$, respectively) in the nonlinear Hall conductivity $\sigma^{xyy}(2\omega;\omega,\omega)$. 
    We set $T = 0.05$ for (a) and $\phi = \pi/3$ for (b). In both of the plots, we set $L=120$ and $\Gamma = 0.05$. 
    }
    \label{fig:phase_temp_dep}
\end{figure}

\textit{Geometric phase and temperature dependence.}---
To examine the nature of divergence in more detail, we focus on the coefficients $\mathcal{A}^{xyy}$ and $\mathcal{B}^{xyy}$, which can be directly calculated from Eq.~\eqref{eq:formula_conductivity} with taking the low-frequency limit~\footnote{The explicit formulas are presented in Supplemental Material~\ref{sec:supp_GF_formula_for_div_factors}.}, and discuss their properties. 
Figure \ref{fig:phase_temp_dep}(a) shows $\mathcal{A}^{xyy}$ and $\mathcal{B}^{xyy}$ for different values of the phase $\phi$. These results clearly show that the AB phase triggers the divergence of the Hall conductivity since these coefficients become nonzero when $\phi \neq 0$ except for several special points, where the time-reversal symmetry is effectively restored, e.g., at $\phi = \pi$. In addition, the coefficients for the two-particle phase have a $\pi$-periodicity with respect to $\phi$, while those for the single-particle phase have a $2\pi$-periodicity. This reflects the fact that the two-particle phase is doubled due to the charge of the electron pairs.

The temperature dependence of $\mathcal{A}^{xyy}$ and $\mathcal{B}^{xyy}$ is presented in Fig.~\ref{fig:phase_temp_dep}(b). We find that the coefficients are finite below the critical temperature, and this clearly shows that the SNHE is unique to the superconducting phase. Whereas $\mathcal{A}^{xyy}$ is suppressed approaching the critical temperature $T_c$, $\mathcal{B}^{xyy}$ is largely enhanced near $T_c$. This will be a useful signature to detect the SNHE experimentally.

\textit{Analysis from the Ginzburg-Landau theory.}---
We show that the SNHE can be understood from the viewpoint of the Ginzburg-Landau (GL) theory, in which the free energy is given by a functional of the complex superconducting order parameter $\psi_\alpha(\bm r)$ ($\alpha$ labels the internal degrees of freedom such as sublattices). Since SNHE is the second-order response in terms of $\bm A$, we need a term of the form 
\begin{equation}
    \sum_{\alpha\beta\mu\nu\lambda} d_{\mu\nu\lambda}^{\alpha\beta}\psi_\alpha^\dagger D_\mu D_\nu D_\lambda \psi_\beta
    \label{eq:Lifshitz invariant}
\end{equation}
in the GL free energy, where $D_\mu=-i\nabla_\mu-e^\ast A_\mu$ ($e^\ast=2$) is the covariant derivative and $d_{\mu\nu\lambda}^{\alpha\beta}$ is a complex coefficient. Here we assume that $d_{\mu\nu\lambda}^{\alpha\beta}$ is symmetric with respect to $\mu, \nu, \lambda$.
The term (\ref{eq:Lifshitz invariant}) is in the odd order of spatial derivatives, and can be seen as a higher-order generalization of the Lifshitz invariant ($d_\mu^{\alpha\beta} \psi_\alpha^\dagger D_\mu \psi_\beta$) \cite{Landau_StatPhys_textbook}. The presence of such a term has been pointed out in the context of, e.g., the stability of second-order phase transitions~\cite{Landau_StatPhys_textbook}, noncentrosymmetric superconductors~\cite{Mineev1994, Mineev2008}, 
parity- and time-reversal symmetry broken superconductors \cite{Kanasugi2022},
and collective modes in superconductors \cite{Kamatani2022, Nagashima2024, Nagashima2024b}.

We require that the third-order Lifshitz invariant (\ref{eq:Lifshitz invariant}) is hermitian, which leads to the condition, $(d_{\mu\nu\lambda}^{\beta\alpha})^\ast=d_{\mu\nu\lambda}^{\alpha\beta}$. We further consider the particle-hole symmetry ($\psi_\alpha\to \psi_\alpha^\dagger$ and $A_\mu \to -A_\mu$), for which we assume that the coefficient transforms as $d_{\mu\nu\lambda}^{\alpha\beta}\to -d_{\mu\nu\lambda}^{\alpha\beta}$
(as expected from the fact that $\mathcal{A}^{xyy}$ is an odd function of the AB flux $\phi$ (Fig.~\ref{fig:phase_temp_dep}(a))). 
Then we find that the coefficient should be symmetric with respect to $\alpha$ and $\beta$ (i.e., $d_{\mu\nu\lambda}^{\beta\alpha}=d_{\mu\nu\lambda}^{\alpha\beta}$), and hence that $d_{\mu\nu\lambda}^{\alpha\beta}$ is real. If the time-reversal symmetry would be present, the coefficient has to satisfy $(d_{\mu\nu\lambda}^{\alpha\beta})^\ast=-d_{\mu\nu\lambda}^{\alpha\beta}$, which is not compatible with real coefficients. Thus the time-reversal symmetry must be broken to have the term (\ref{eq:Lifshitz invariant}).
One might wonder if the inversion symmetry should also be broken to have the odd derivative term (\ref{eq:Lifshitz invariant}). However, this is not necessarily the case: If $\psi_\alpha$ belongs to a nontrivial representation under inversion ($D_\mu \to -D_\mu$), the term (\ref{eq:Lifshitz invariant}) can exist. For example, when $\psi_\alpha$ has two components ($\alpha,\beta=1,2$) and transforms as $\psi_1\to \psi_2$ and $\psi_2\to\psi_1$ under inversion, the term (\ref{eq:Lifshitz invariant}) is allowed to exist if $d_{\mu\nu\lambda}^{11}=-d_{\mu\nu\lambda}^{22}$ and $d_{\mu\nu\lambda}^{12}=d_{\mu\nu\lambda}^{21}=0$. 

Further conditions are imposed if one considers the point group symmetry.
Let us denote the representation of $\psi_\alpha$ and $D_j$ as $[\psi_\alpha]$ and $[D_\mu]$.
In our model, the point group is $C_{3h}$ (in which the inversion symmetry is broken), and the order parameter ($\psi_1\sim\Delta_A$ and $\psi_2\sim\Delta_B$) belongs to the trivial representation ($[\psi_\alpha]=A'$). With $[D_\mu]=A''\oplus E'$, we obtain
\begin{equation}
    [\psi_\alpha^\dagger] \!\otimes\! [D_\mu] \!\otimes\! [D_\nu] \!\otimes\! [D_\lambda] \!\otimes\! [\psi_\beta]
    \!=\!
    2A'\oplus 7A'' \oplus 6E' \oplus 3E'',
\end{equation}
which contains the trivial representation ($A'$). Thus the third-order Lifshitz invariant (\ref{eq:Lifshitz invariant}) can exist in our model from the symmetry point of view. This is in sharp contrast to the case of the first-order Lifshitz invariant, which is not allowed to exist in our model according to the general classification of the Lifshitz invariant~\cite{Nagashima2024}.

Now the SNHE can be understood in the GL theory as follows. The current is given by the derivative of the free energy density, $j^\mu=-\partial f/\partial A_\mu$. If the third-order Lifshitz invariant (\ref{eq:Lifshitz invariant}) is present, the current in the second order of $\bm E$ is given by
\begin{equation}
    j^{\mu(2)}(\omega)
    \propto
    \frac{1}{\omega^2} \sum_{\alpha\beta \nu \lambda} d_{\mu \nu \lambda}^{\alpha\beta} \psi_\alpha^\dagger \psi_\beta E_\nu(\omega) E_\lambda(\omega).    
\end{equation}
Thus the nonlinear Hall conductivity becomes $\sigma^{xyy}(2\omega; \omega, \omega)\propto \frac{1}{\omega^2} \sum_{\alpha\beta} d_{xyy}^{\alpha\beta} \psi_\alpha^\dagger \psi_\beta$,
which generally diverges as $\frac{1}{\omega^2}$ in the dc limit.
If the inversion symmetry were present in a system with a two-component order parameter, the coefficient of $\frac{1}{\omega^2}$ would be given by $d_{xyy}^{11}(\psi_1^\dagger \psi_1-\psi_2^\dagger \psi_2)$. If $|\psi_1|=|\psi_2|$ (i.e., the inversion symmetry is not spontaneously broken), the SNHE vanishes. Therefore, the inversion symmetry must be either explicitly or spontaneously broken for the system to exhibit the SNHE.
Since the GL argument is solely based on symmetries, the SNHE can be observed in a wide range of superconductors without inversion and time-reversal symmetries, not limited to the microscopic model employed in this study.

\begin{figure}[t]
    \includegraphics[width=8.8cm]{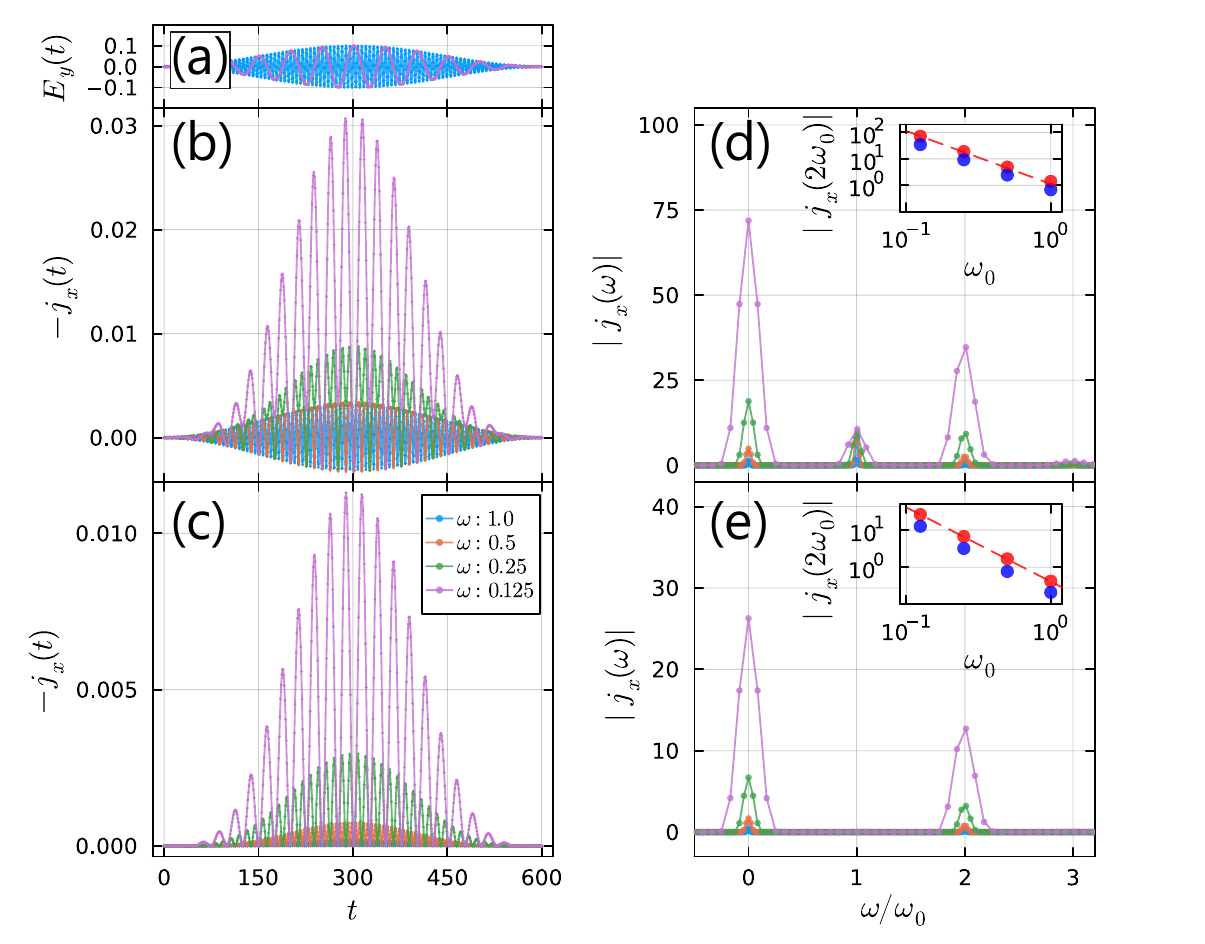}
    \caption{
        Real-time dynamics driven by a light field $(0, E_y(t))$ where $E_y(t) =  (1-\cos(2\pi t/\tau)) \cos(\omega_0 t)$ with $\tau = 300$, whose wave forms are shown in (a) for $\omega_0 = 1.0, 0.125$.
        (b, c) The Hall current $j^y(t)$. (d, e) Fourier spectrum of the Hall current $|j^y(\omega)|$. The inset shows the $\omega_0$-dependence of the peak heights at $\omega=0$ (red) and $\omega=2\omega_0$ (blue) in the Fourier spectrum. The red dashed lines are proportional to $1/\omega_0^2$. (b) and (d) [(c) and (e)] are for \textsf{SC1} [\textsf{SC2}]. We set $\phi=\pi/3$ and $L=24$. The initial temperature is set to $T=0.05$.}
    \label{fig:light_drive}
\end{figure}

\textit{Light-driven real-time dynamics. }--- 
Finally, we clarify the connection between the SNHE and light-driven dynamics by performing a real-time simulation.
We numerically solve the equation of motion obtained from the mean-field approximation of the Hamiltonian $\hat{H}_\mathrm{sys}$, and track the Hall current $j^y(t)$ in time under the light pulse $\bm E(t) = (0, E_y(t))$ with $E_y(t) =  (1-\cos(2\pi t/\tau)) \cos(\omega t)$. Note that this calculation does not contain the effect of heat baths. The results are shown in Figs.~\ref{fig:light_drive}(b) and (c), where we find that the Hall current is significantly rectified in the low-frequency regime. This is a direct consequence of the SNHE. In fact, the peak heights of the dc and $2\omega$ components show the divergent behavior in the low-frequency limit,
as shown in Figs.~\ref{fig:light_drive}(d) and (e). This calculation fully contains the effect of collective modes, and thus shows the robustness of the SNHE against the existence of the collective modes.

\textit{Experimental setup.}---
The key ingredient of the SNHE studied in this paper is the geometric phase $\phi$. One way for the experimental realization is via the Floquet engineering \cite{Bukov2015, Oka2019, Weitenberg2021}: Application of circularly polarized light can effectively realize the Haldane model~\cite{Oka2009}, which has been confirmed in ultracold atoms~\cite{Jotzu2014} and electron systems~\cite{McIver2020}. Thus, this is a promising approach to realize the term in Eq.~\eqref{eq:H1}. The two-particle one is also effectively realizable with the Floquet engineering, while there appear additional terms~\footnote{We provide the derivation for pair hopping terms similar to ones in Eq.~\eqref{eq:H2} based on the Floquet theory in Supplemental Material \ref{sec:supp_Floquet}.}. The other is via the loop current order: One can often model the loop current order in the mean-field theory
by the geometric phase~\cite{Varma2006, Watanabe2021}, and thus the coexistence of superconductivity and loop current order can be a platform for the SNHE. Furthermore, a recent study has discussed the loop \textit{supercurrent} order~\cite{Ghosh2021}, which may realize the two-particle AB phase. For measurements, we can choose various approaches used for Hall responses, including transport measurements and optical spectroscopy. To combine with the Floquet engineering approach, the pump-probe optical measurement is promising, allowing us to detect the real-time dynamics of rectified current shown in Fig.~\ref{fig:light_drive}(b).

\textit{Conclusion.}--- 
We have proposed a mechanism for the superconducting nonlinear Hall effect driven by geometric phases, and have demonstrated the robust divergence of the nonlinear Hall conductivity in the dc limit in $s$-wave superconductors. We have also discussed the SNHE from the viewpoint of the Ginzburg-Landau theory, and have shown the connection to light-driven dynamics. 

\textit{Acknowledgments.}--- 
    We would like to thank Akito Daido, Taisei Kitamura, Hiroto Tanaka, Hikaru Watanabe, Youichi Yanase and Huanyu Zhang for helpful discussions. This work is supported by JST FOREST (Grant No.~JPMJFR2131), JST PRESTO (Grant No.~JPMJPR2256) and JSPS KAKENHI (Grant Nos.~JP22K20350, JP23K17664, and JP24H00191).

\bibliography{ref.bib}

\clearpage


\newcommand{\supplementary}{
  \setcounter{section}{0}
  \renewcommand{\thesection}{S\arabic{section}}
  \setcounter{equation}{0}
  \renewcommand{\theequation}{S\arabic{equation}}
  \setcounter{figure}{0}
  \renewcommand{\thefigure}{S\arabic{figure}}
  \setcounter{table}{0}
  \renewcommand{\thetable}{S\arabic{table}}
  \setcounter{page}{1}
  \renewcommand{\thepage}{\arabic{page}}
}

\supplementary
\onecolumngrid

\begin{center}
    \textbf{Supplemental Material for ``Superconducting nonlinear Hall effect induced by geometric phases"} \vspace{3mm}  \\
    Kazuaki Takasan$^1$ and Naoto Tsuji$^{1,2}$
    \vspace{1mm}\\
$^1$\textit{Department of Physics, University of Tokyo, 7-3-1 Hongo, Tokyo 113-0033, Japan} \\
$^2$\textit{RIKEN Center for Emergent Matter Science (CEMS), 2-1 Hirosawa, Wako, Saitama 351-0198, Japan}
\end{center}

\vspace{5mm}

\twocolumngrid

\section{Outline of the derivation of the formula for the nonlinear conductivity}
\label{sec:supp_derivation_nonlinear_cond}

In this section, we outline the derivation of the formula for the second-order nonlinear conductivity, Eq.~\eqref{eq:formula_conductivity}, in the main text.
The second-order nonlinear conductivity is defined as
\begin{equation}
    j^{\mu(2)}(t) \!=\! \sum_{\nu \lambda} \int \!\!\! dt' \!\! \int \!\!\! dt'' \sigma^{\mu \nu \lambda}(t-t', t-t'') E_\nu(t') E_\lambda(t''),
\end{equation}
where $j^{\mu(2)}(t)$ is the second-order current density with respect to the applied electric field $E_\mu(t)$. By performing the Fourier transformation, we obtain
\begin{align}
    j^{\mu(2)}(\omega) &\!=\! \sum_{\nu \lambda}  \int \!\! \frac{d\omega_1}{2\pi} \!\! \int \!\! \frac{d\omega_2}{2\pi} \sigma^{\mu \nu \lambda}(\omega_{12}; \omega_1, \omega_2) \nonumber \\
    &\qquad \qquad \times  2\pi \delta(\omega - \omega_{12})  E_\nu(\omega_1) E_\lambda(\omega_2). \label{eq:Fourier_current}
\end{align}
Our goal is to derive the formula for $\sigma^{\mu \nu \lambda}(\omega_{12}; \omega_1, \omega_2)$.

For this purpose, let us consider the current expectation value, 
\begin{equation}
    J^\mu(t) 
    = - \left\langle \frac{\partial \hat{H}(t)}{\partial A_\mu(t)} \right\rangle,
\end{equation}
where
\begin{equation}
    \hat{H}(t)
    = \sum_{\bm k} \hat{\bm \psi}^\dagger_{\bm k} h^\mathrm{BdG}_{\bm k, \bm A(t)} \hat{\bm \psi}_{\bm k} + \mathcal{E}_{0, \bm A(t)}, 
\end{equation}
$\hat{\bm \psi}_{\bm k} = (\hat{\psi}_{\bm k, 1}, \hat{\psi}_{\bm k, 2}, \hat{\psi}_{\bm k, 3}, \hat{\psi}_{\bm k, 4})  = (\hat{a}_{\bm k, \uparrow}, \hat{b}_{\bm k, \uparrow}, \hat{a}_{-\bm k, \downarrow}^\dagger, \hat{b}_{-\bm k, \downarrow}^\dagger)^\mathsf{T}$ [$\hat{a}_{\bm k, \sigma}$($\hat{b}_{\bm k, \sigma}$) is the annihilation operator for electrons with momentum $\bm k$ and spin $\sigma$ in the sublattice $\mathbb{A}$($\mathbb{B}$)], $h^\mathrm{BdG}_{\bm k, \bm A(t)}$ is the Bogoliubov-de Gennes Hamiltonian, and $\mathcal{E}_{0, \bm A(t)}$ is the energy shift coming with the mean-field approximation (see the main text for their explicit forms). The current expectation value is expressed as
\begin{equation}
    J^\mu(t) 
    = i \sum_{\bm k } \mathrm{Tr} [V^\mu_{\bm k, \bm A(t)} G^{<}_{\bm k, \bm A(t) }(t, t)] + J_{0, \bm A(t)}^\mu, \label{eq:current_Green_func}   
\end{equation}
where $V^\mu_{\bm k, \bm A(t)} = \frac{\partial h^\mathrm{BdG}_{\bm k, \bm A(t)}}{\partial A_\mu(t)}|_{A=0}$, $G^{<}_{\bm k, \bm A(t)}(t, t')$ is the lesser Green's function defined as $[G^{<}_{\bm k, \bm A(t)}(t, t')]_{ij}= i \langle \hat{\psi}_{\bm k, j}(t) \hat{\psi}^\dagger_{\bm k, i}(t') \rangle$~\cite{Kamenev_book}, and $J_{0, \bm A(t)}^\mu= - \frac{\partial \mathcal{E}_{0, \bm A(t)}}{\partial A_\mu(t)}$. To obtain the nonlinear conductivity, we expand the current expectation value [Eq.~\eqref{eq:current_Green_func}] in terms of the vector potential $\bm A$. 

The expansion of $V^\mu_{\bm k, \bm A}$ and $J_{0, \bm A}^\mu$ can be carried out as below:
\begin{gather}
    V^\mu_{\bm k, \bm A} \!=\! V^{\mu (0)}_{\bm k} \!+\! V^{\mu (1)}_{\bm k} \!+\! V^{\mu (2)}_{\bm k} \!\!+\! \mathcal{O}(A^3),~ V^{\mu (0)}_{\bm k} = V^\mu_{\bm k}, \nonumber\\
    \quad V^{\mu (1)}_{\bm k} = \sum_{\nu} W^{\mu \nu}_{\bm k} A_\nu,~V^{\mu (2)}_{\bm k} = \frac{1}{2} \sum_{\nu \lambda} X^{\mu\nu\lambda}_{\bm k} A_\nu A_\lambda, \\
    J_{0, \bm A}^\mu = -\mathcal{E}_0^\mu \!-\! \sum_{\nu} \mathcal{E}_0^{\mu \nu}\! A_\nu \!-\! \frac{1}{2} \sum_{\nu \lambda} \mathcal{E}_0^{\mu\nu\lambda} \! A_\nu A_\lambda \!+\! \mathcal{O}(A^3), 
\end{gather}
where $V_{\bm k}^\mu = V_{\bm k, \bm A}|_{A=0}$, $W^{\mu \nu}_{\bm k} = \frac{\partial V^{\mu}_{\bm k, \bm A}}{\partial A_\nu}|_{A=0}$, $X^{\mu\nu\lambda}_{\bm k} = \frac{\partial^2 V^{\mu}_{\bm k, \bm A}}{ \partial A_\nu \partial A_\lambda} $, $\mathcal{E}_0^\mu = \frac{\partial \mathcal{E}_{0, \bm A}}{\partial A_\mu}|_{A=0}$, $\mathcal{E}_0^{\mu \nu} = \frac{\partial^2 \mathcal{E}_{0, \bm A}}{\partial A_\mu \partial A_\nu}|_{A=0}$, and $\mathcal{E}_0^{\mu\nu\lambda} = \frac{\partial^3 \mathcal{E}_{0, \bm A}}{\partial A_\mu \partial A_\nu \partial A_\lambda}|_{A=0}$.
To obtain the expansion of $G^{<}_{\bm k, \bm A}$,
\begin{equation}
    G_{\bm k, \bm A}^< = G_{\bm k}^{<(0)} + G_{\bm k}^{<(1)} + G_{\bm k}^{<(2)} + \mathcal{O}(A^3),
\end{equation}
we use the Keldysh equation for the lesser Green's function \cite{Kamenev_book},
\begin{align}
    G^{<}_{\bm k, \bm A} &= G^{R}_{\bm k, \bm A} * \Sigma^{<} * G^{A}_{\bm k, \bm A}. ~\label{eq:Keldysh_eq}
\end{align}
Sec.~\ref{sec:supp_bath} for more details.
Here, in order to simplify the notation, we have omitted the time arguments and introduced the convolution product defined as 
\begin{equation}
    [A * B ](t, t') = \int_{-\infty}^{\infty} \!\! dt'' A(t, t'') B(t'', t').
\end{equation}
If either $A$ or $B$ is a function with a single time argument such as $C(t)$, we use $\tilde{C}(t, t') = C(t) \delta(t-t')$ instead of $C(t)$. The self-energy $\Sigma$ is introduced to describe the effect of dissipation induced by the heat bath. In particular, the lesser component $\Sigma^<$ gives information about the distribution of electrons in the bath. The explicit form will be discussed in Eq.~\eqref{eq:lesser_self_energy} in Sec.~\ref{sec:supp_bath}.

We put the expansion of the retarded and advanced Green's functions,
\begin{equation}
    G_{\bm k, \bm A}^{R/A} = G_{\bm k}^{R/A(0)} + G_{\bm k}^{R/A(1)} + G_{\bm k}^{R/A(2)} + \mathcal{O}(A^3),
\end{equation}
with
\begin{align}
    G_{\bm k }^{R/A(1)} &\!=\! G_{\bm k }^{R/A(0)} \!*\! (\bm V_{\bm k} \cdot \bm A) \!*\! G_{\bm k }^{R/A(0)}, \\
    G_{\bm k }^{R/A(2)} &\!=\! G_{\bm k }^{R/A(0)} \!*\! (\bm V_{\bm k} \cdot \bm A) \!*\! G_{\bm k }^{R/A(0)}\!*\! (\bm V_{\bm k} \cdot \bm A)  \!*\! G_{\bm k }^{R/A(0)} \nonumber \\
    &\quad + \frac{1}{2} G_{\bm k }^{R/A(0)} \!*\! ( \bm A^T W_{\bm k} \bm A) \!*\! G_{\bm k }^{R/A(0)},
\end{align}
into Eq.~\eqref{eq:Keldysh_eq}, and then obtain
\begin{align}
    G^{<(0)}_{\bm k } &= G^{R(0)}_{\bm k } * \Sigma^{<} * G^{A(0)}_{\bm k },\\
    G^{<(1)}_{\bm k } &= G^{R(1)}_{\bm k } * \Sigma^{<} * G^{A(0)}_{\bm k }+ G^{R(0)}_{\bm k } * \Sigma^{<} * G^{A(1)}_{\bm k },\\
    G^{<(2)}_{\bm k } &= G^{R(2)}_{\bm k } * \Sigma^{<} * G^{A(0)}_{\bm k } + G^{R(1)}_{\bm k } * \Sigma^{<} * G^{A(1)}_{\bm k } \nonumber \\
    &\quad ~ + G^{R(0)}_{\bm k } * \Sigma^{<}* G^{A(2)}_{\bm k },
\end{align}
where the Green's function $G^{R/A(0)}_{\bm k}$ is obtained from the Dyson's equation,
\begin{equation}
    [i \partial_{t} - h_{\bm k, \bm 0}^\text{BdG} - \Sigma^{R/A}]G^{R/A(0)}_{\bm k}(t, t') = \delta(t-t').
\end{equation}
Note that the effect of dissipation is taken into account through the retarted/advanced component of the self-energy. Their explicit forms are derived in Eqs.~\eqref{eq:retarted_self_energy} and \eqref{eq:advanced_self_energy} in Sec.~\ref{sec:supp_bath}. 

Combining all the expansions, we reach
\begin{align}
    J^{\mu(2)}(t)&= 
    i \sum_{\bm k } \mathrm{Tr} \left[ V^{\mu (0)}_{\bm k}(t)G^{<(2)}_{\bm k }(t, t) \right] \nonumber \\
    & \quad +i \sum_{\bm k } \mathrm{Tr} \left[ V^{\mu (1)}_{\bm k}(t)G^{<(1)}_{\bm k }(t, t) \right]  \nonumber
    \\
    & \qquad +i \sum_{\bm k } \mathrm{Tr} \left[ V^{\mu (2)}_{\bm k}(t)G^{<(0)}_{\bm k }(t, t) \right] \nonumber \\
    &\quad \qquad - \frac{1}{2} \sum_{\nu \lambda} \mathcal{E}_0^{\mu\nu\lambda} \! A_\nu(t) A_\lambda(t). \label{eq:current_expansion}
\end{align}
Putting all the explicit form of the expansion coefficients, simplifying the expression, and performing the Fourier transformation, we arrive at the form of Eq.~\eqref{eq:Fourier_current}. Then, we can extract the formula for the second-order nonlinear conductivity, which is Eq.~\eqref{eq:formula_conductivity} in the main text. In the simplification, we use a relation,
\begin{equation}
    G^{R(0)}_{\bm k } * \Sigma^{<}_{\bm k } * G^{A(0)}_{\bm k } = f * (G^{A(0)}_{\bm k }  - G^{R(0)}_{\bm k })
\end{equation}
where $f(t)$ is the Fourier transform of the Fermi distribution function at temperature $T$. 

\twocolumngrid

\section{Self-energy for a fermionic bath}
\label{sec:supp_bath}

In this section, we derive the self-energy that represents the effect of a fermionic bath used in the derivation of Eq.~\eqref{eq:formula_conductivity} outlined in Sec.~\ref{sec:supp_derivation_nonlinear_cond}.  

We consider a fermionic lattice model coupled to a fermionic bath,
\begin{equation}
    \hat{H}_\mathrm{tot} = \hat{H}_\mathrm{sys} + \hat{H}_\mathrm{bath} + \hat{H}_\mathrm{int},    
\end{equation}
where
\begin{gather*}
    \hat{H}_\mathrm{sys} = \hat{H}_\text{free} + \hat{H}',~~~ \hat{H}_\text{free}= \sum_{i,j = 1}^N \sum_{\sigma=\uparrow, \downarrow} t_{ij} \hat{c}_{i\sigma}^\dagger \hat{c}_{j\sigma}, \\
    \hat{H}_\mathrm{bath} = \sum_{i=1}^N \sum_{m=1}^M \sum_{\sigma=\uparrow, \downarrow} \varepsilon_m \hat{b}_{i m \sigma}^\dagger \hat{b}_{i m \sigma},  \\
    \hat{H}_\mathrm{int} =  \sum_{i=1}^N \sum_{m=1}^M \sum_{\sigma=\uparrow, \downarrow} V_{m} \hat{c}_{i\sigma}^\dagger \hat{b}_{im\sigma} + \mathrm{h.c.} 
\end{gather*}
Here, $\hat{c}_{i\sigma}$ ($\hat{b}_{im\sigma}$) is the annihilation operator for electrons in the system (bath) at site $i$ with spin $\sigma$, $t_{ij}$ is the hopping amplitude, $\varepsilon_m$ is the energy level of the bath, and $V_m$ is the coupling strength between the system and the bath. $\hat{H}'$ represents the remaining terms in the system which are not written in the quadratic form, e.g., interaction terms. $N$ ($M$) is the number of sites (modes) in the system (bath). $M$ fermion modes are located at each site, coupling locally to the system. This type of bath is often called the B\"uttiker bath, is and used in various studies. This bath can be interpreted as a metallic substrate coupled to the system.

Let us move to the momentum space by introducing the Fourier transformation. For $\hat{H}_\text{free}$, we obtain
\begin{equation}
    \hat{H}_\text{free} = \sum_{\bm k} \sum_{n=1}^{N_\text{b}} \sum_{\sigma=\uparrow, \downarrow} e_{\bm k n} \hat{c}_{\bm k n \sigma}^\dagger \hat{c}_{\bm k n \sigma},
\end{equation}
where $e_{\bm k n}$ is the energy dispersion of the system. Here, we have assumed the number of bands in the system is $N_\text{b}$. For the honeycomb lattice model used in the main text, $N_\text{b}=2$. In the same way, we obtain the bath Hamiltonian and the interaction Hamiltonian in the momentum space,
\begin{gather}
    \hat{H}_\mathrm{bath} = \sum_{\bm k} \sum_{n=1}^{N_\text{b}} \sum_{m=1}^M \sum_{\sigma=\uparrow, \downarrow} \varepsilon_{m} \hat{b}_{\bm k n m \sigma}^\dagger \hat{b}_{\bm k n m \sigma}, \\
    \hat{H}_\mathrm{int} = \sum_{\bm k} \sum_{n=1}^{N_\text{b}} \sum_{m=1}^M \sum_{\sigma=\uparrow, \downarrow}  \sum_{\bm k} V_{m} c_{\bm k n \sigma}^\dagger b_{\bm k n m\sigma} + \mathrm{h.c.}.
\end{gather}

To consider the superconducting state, we introduce the Nambu basis for both the system and the bath:
$( \hat{\psi}^\dagger_{\bm k n 1}, \hat{\psi}^\dagger_{\bm k n 2} ) = ( \hat{c}^\dagger_{\bm k n \uparrow}, \hat{c}_{- \bm k n \downarrow} )$ and $(\hat{\phi}_{\bm k n m1}, \hat{\phi}_{\bm k n m2}) = (\hat{b}_{\bm k n m \uparrow}, \hat{b}_{- \bm k n m \downarrow}^\dagger)$, respectively. Using these operators, we can rewrite the Hamiltonian as
\begin{gather}
    \hat{H}_\mathrm{bath} = \sum_{\bm k} \sum_{n=1}^{N_\text{b}} \sum_{m=1}^M \sum_{l=1,2} E_{ml} \hat{\phi}_{\bm k n m l}^\dagger \hat{\phi}_{\bm k n m l}
    + \text{const.}  \\
    H_\mathrm{int}
    = \sum_{\bm k} \sum_{n=1}^{N_\text{b}} \sum_{m=1}^M \sum_{l=1,2} \left( V_{ml} \hat{\psi}_{\bm k n l}^\dagger \hat{\phi}_{\bm k n m l} + \mathrm{h.c.} \right),
\end{gather}
where $(E_{m1}, E_{m2})=(\varepsilon_{m},-\varepsilon_{m})$ and $(V_{m1},V_{m2}) = (V_m, - V_m^*)$.

To obtain the self-energy easily, we move to the path-integral representation. The total action $S_\mathrm{tot}$ is given by 
\begin{align*}
    &S_\mathrm{tot}[\psi, \bar{\psi}, \phi, \bar{\phi}]\\
    &\!= S_\mathrm{sys}[\psi, \bar{\psi}] \!+\!\! \int_C\!\!dt \sum_{\bm k n m} \sum_{l=1,2} [ \bar{\phi}_{\bm k n m l}(t) \left( i\partial_t - E_{ml} \right) \phi_{\bm k n ml}(t) \\
    &\qquad \qquad \qquad - V_{ml} \bar{\psi}_{\bm k n l}(t) \phi_{\bm k n m l}(t) - V^*_{ml} \bar{\phi}_{\bm k n m l}(t) \psi_{\bm k n l}(t)],
\end{align*}
where $\phi_{\bm k n m l}(t)$ and $\psi_{\bm k n l}(t)$ are the Grassmann variables for the bath and the system, respectively, and $\bar{\phi}_{\bm k n m l}(t)$ and $\bar{\psi}_{\bm k n l}(t)$ are their conjugate variables. $S_\mathrm{sys}[\psi, \bar{\psi}]$ is the action for the system. $C$ denotes the Keldysh contour~\cite{Kamenev_book}. Integrating out the bath degree of freedom, we obtain the effective action for the system,
\begin{align*}
    S_\mathrm{eff}[\psi, \bar{\psi}]
    &= S_\mathrm{sys}[\psi, \bar{\psi}] \\
    &\quad - \int_C\!\!dt \int_C\!\!dt' \sum_{\bm k} \bar{\bm \psi}_{\bm k}(t)\hat{\Sigma}(t, t') \bm \psi_{\bm k}(t'),
\end{align*}
where $\bar{\bm \psi}_{\bm k} = (\bar{\psi}_{\bm k 1 1}, \bar{\psi}_{\bm k 2 1}, \cdots, \bar{\psi}_{\bm k N_\text{b} 1}, \bar{\psi}_{\bm k 1 2}, \bar{\psi}_{\bm k 2 2}, \cdots, \bar{\psi}_{\bm k N_\text{b} 2})$ and $\bm \psi_{\bm k} = (\psi_{\bm k 1 1}, \psi_{\bm k 2 1}, \cdots, \psi_{\bm k N_\text{b} 1}, \psi_{\bm k 1 2}, \psi_{\bm k 2 2}, \cdots, \psi_{\bm k N_\text{b} 2})^\mathsf{T}$. The self-energy $\Sigma(t, t')$ is given by
\begin{equation}
    \Sigma(t, t') =
    \sum_m 
    \begin{pmatrix}
        |V_{m}|^2  g_{m1}(t, t') 1_{N_\text{b}} & \\
        & |V_{m}|^2  g_{m2}(t, t') 1_{N_\text{b}}
    \end{pmatrix},
\end{equation}
where $g_{ml}(t, t')$ is the Green's function for the bath defined as $[g_{ml}(t, t')]^{-1} = (i\partial_t - E_{ml})$ and $1_{N_\text{b}}$ is the $N_\text{b} \times N_\text{b}$ unit matrix. Moving to the frequency space and taking the retarted components, we obtain
\begin{equation}
    \Sigma^R(\omega) =
    \sum_m 
    \begin{pmatrix}
        \displaystyle \frac{|V_{m}|^2}{\omega - \varepsilon_{m} + i0} 1_{N_\text{b}} & \\
        & \displaystyle \frac{|V_{m}|^2}{\omega + \varepsilon_{m} + i0} 1_{N_\text{b}}
    \end{pmatrix}.  
\end{equation}
Using a formula
$\frac{1}{\omega+i0^+} = \mathcal{P}\frac{1}{\omega} - i \pi \delta(\omega)$, where $\mathcal{P}$ is the principal value, we obtain
\begin{equation}
    \frac{|V_{m}|^2}{\omega \mp \varepsilon_{m} + i0} = \sum_m \mathcal{P}\frac{|V_{m}|^2}{\omega \mp \varepsilon_{m}} - i \Gamma(\pm \omega) 
\end{equation}
with $\Gamma(\omega) = \pi \sum_m |V_{m}|^2 \delta(\omega - \varepsilon_m)$. The real part only gives the energy shift that can be canceled out by redefining the chemical potential of the system, and thus we omit the energy shift. Also, we assume the flat density of states in the bath, and then $\Gamma(\omega)$ is independent of $\omega$ ($\Gamma(\omega)=\Gamma$). Finally, we obtain the self-energy as
\begin{equation}
    \Sigma^R(\omega) \simeq - i \Gamma \cdot 1_{2N_\text{b}}. \label{eq:retarted_self_energy}
\end{equation}
In the same way, the advanced component can be obtained as 
\begin{equation}
    \Sigma^A(\omega) \simeq i \Gamma \cdot 1_{2N_\text{b}}. \label{eq:advanced_self_energy}
\end{equation}
For the lesser component, we obtain
\begin{equation}
    \Sigma^<(\omega) = 2 i f(\omega)
    \begin{pmatrix}
        \Gamma(\omega) 1_{N_\text{b}} & \\
        & \Gamma(-\omega) 1_{N_\text{b}}
    \end{pmatrix},
\end{equation}
where $f(\omega)$ is the Fermi distribution function with temperature $T$. Assuming the flat density of states in the bath as before, we obtain
\begin{equation}
    \Sigma^<(\omega) \simeq 2 i \Gamma f(\omega) \cdot 1_{2N_\text{b}}. \label{eq:lesser_self_energy}
\end{equation}
The explicit forms Eqs.\eqref{eq:retarted_self_energy}, \eqref{eq:advanced_self_energy}, and \eqref{eq:lesser_self_energy} are used in Sec.~\ref{sec:supp_derivation_nonlinear_cond}.

\section{Derivation of an analytic formula for the nonlinear Hall conductivity}
\label{sec:supp_ana_derivation}

In this section, we derive the analytic formula of the second-order Hall conductivity arising from the pair-hopping terms. Note that we do not consider the effect of dissipation in this derivation, but the divergent behavior is consistent with the results with dissipation.

As mentioned in Sec.~\ref{sec:supp_derivation_nonlinear_cond}, the current expectation value is given by 
\begin{equation}
    J^\mu(t) 
    = - \left\langle \frac{\partial \hat{H}(t)}{\partial A_\mu(t)} \right\rangle,
\end{equation}
where
\begin{equation}
    \hat{H}(t)
    = \sum_{\bm k} \hat{\bm \psi}^\dagger_{\bm k} h^\mathrm{BdG}_{\bm k, \bm A(t)} \hat{\bm \psi}_{\bm k} + \mathcal{E}_{0, \bm A(t)}, 
\end{equation}
$\hat{\bm \psi}_{\bm k} = (\hat{\psi}_{\bm k, 1}, \hat{\psi}_{\bm k, 2}, \hat{\psi}_{\bm k, 3}, \hat{\psi}_{\bm k, 4})  = (\hat{a}_{\bm k, \uparrow}, \hat{b}_{\bm k, \uparrow}, \hat{a}_{-\bm k, \downarrow}^\dagger, \hat{b}_{-\bm k, \downarrow}^\dagger)^\mathsf{T}$ [$\hat{a}_{\bm k, \sigma}$($\hat{b}_{\bm k, \sigma}$) is the annihilation operator for electrons with momentum $\bm k$ and spin $\sigma$ in the sublattice $\mathbb{A}$($\mathbb{B}$)], $h^\mathrm{BdG}_{\bm k, \bm A(t)}$ is the Bogoliubov-de Gennes Hamiltonian, and $\mathcal{E}_{0, \bm A(t)}$ is the energy shift coming with the mean-field approximation. 

Differently from Sec.~\ref{sec:supp_derivation_nonlinear_cond}, we will evaluate the current expectation value itself. Then, we will expand it in terms of the electric field and obtain the formula for the second-order nonlinear conductivity. The expectation value of current density is expressed as
\begin{equation}
    j^\mu_{\bm A} = \frac{1}{N_\mathrm{s}}\sum_{\bm k} \hat{\bm \psi}^\dagger_{\bm k}     
    \begin{pmatrix}
        - v_{\mathrm{s}, \bm k - \bm A}^\mu & v_{\mathrm{p}, \bm A}^\mu \\
        [v_{\mathrm{p}, \bm A}^{\mu}]^\dagger & [v_{\mathrm{s}, -\bm k - \bm A}^{\mu}]^\mathsf{T}
    \end{pmatrix} \hat{\bm \psi}_{\bm k}
     +  \frac{1}{N_\mathrm{s}}\frac{\partial \mathcal{E}_{0, \bm A}}{\partial A_\mu}, 
\end{equation}
where
\begin{gather}
    v_{\mathrm{s}, \bm k}^{\mu} = \frac{\partial h^\mathrm{N}_{\bm k}}{\partial k_\mu} 
    =  
    \begin{pmatrix}
        t_2 g_{2}^\mu(\bm k, \phi)& t_1 g_{1}^\mu(\bm k) \\
        t_1 g_{1}^\mu(\bm k)^* & t_2 g_{2}^\mu(\bm k, -\phi)
    \end{pmatrix},
    \\
    v_{\mathrm{p}, \bm A}^\mu = \frac{\partial \Delta_{\bm A}}{\partial A_\mu}=
    \begin{pmatrix}
        v_{\mathrm{p}1, \bm A}^{\mu} & \\
        &v_{\mathrm{p}2, \bm A}^{\mu}
    \end{pmatrix},\\
    v_{\mathrm{p}1, \bm A}^{\mu} = \frac{2J_2}{U} g_2^\mu(-2\bm A, 2 \phi) \Delta_{\mathrm{A}} + \frac{2J_1}{U} g_1^\mu(-2\bm A) \Delta_{\mathrm{B}},\\
    v_{\mathrm{p}2, \bm A}^{\mu}  = \frac{2J_2}{U} g_2^\mu(-2\bm A, -2 \phi) \Delta_{\mathrm{B}} + \frac{2J_1}{U} (g_1^\mu(-2\bm A))^* \Delta_{\mathrm{A}},
\end{gather}
$g_{1}^\mu(\bm k) = \partial_{k_\mu} g_1(\bm k)$, $g_{2}^\mu(\bm k, \phi) = \partial_{k_\mu} g_2(\bm k, \phi)$, and $N_{\mathrm{s}}$ is the number of sublattices in the system ($N_\mathrm{s}$ is half of the total number of lattice sites). 

Let us divide the current expectation value as
\begin{equation}
    j^\mu_{\bm A} = j_{\mathrm{s}, \bm A}^\mu + j_{\mathrm{p}, \bm A}^\mu,
\end{equation}
where $j_{\mathrm{s}, \bm A}^\mu$ and $j_{\mathrm{p}, \bm A}^\mu$ are the contributions from the single-particle hopping terms ($\propto$ $t_1$ or $t_2$) and the pair hopping terms ($\propto$ $J_1$ or $J_2$), respectively. $j_{\mathrm{p}, \bm A}^\mu$ is given by
\begin{align}
    j_{\mathrm{p}, \bm A}^\mu 
    &\!=\! \frac{1}{N_\mathrm{s}}
    \sum_{\bm k} \left\{ 
        v_{\mathrm{p}1, \bm A}^{\mu} 
        \langle a_{\bm k \uparrow}^\dagger  a_{-\bm k \downarrow}^\dagger \rangle 
        \!+\! v_{\mathrm{p}2, \bm A}^{\mu} \langle  b_{\bm k \uparrow}^\dagger  b_{-\bm k \downarrow}^\dagger \rangle \right. \nonumber
            \\
        & \left. \quad \!\!\!
            +(v_{\mathrm{p}1, \bm A}^\mu)^*
            \langle a_{-\bm k \downarrow}   a_{\bm k \uparrow} \rangle 
            \!+\!(v_{\mathrm{p}2, \bm A}^{\mu})^*
            \langle  b_{-\bm k \downarrow} b_{\bm k \uparrow} \rangle
    \right\}
    \!+\!  \frac{1}{N_\mathrm{s}}\frac{\partial \mathcal{E}_{0, \bm A}}{\partial A_\alpha}.
\end{align}
Since $v_{\mathrm{p}1, \bm A}^\mu$ and $v_{\mathrm{p}2, \bm A}^\mu$ do not depend on $\bm k$, we can perform the momentum summation. The expectation values are given by
\begin{align}
    j_{\mathrm{p}, \bm A}^\mu &=  
        -\frac{v_{\mathrm{p}1, \bm A}^{\mu}}{U} 
            \Delta^*_\mathrm{A}
        -\frac{v_{\mathrm{p}2, \bm A}^{\mu}}{U}   
            \Delta^*_\mathrm{B} \nonumber \\
            &\quad 
        -\frac{(v_{\mathrm{p}1, \bm A}^\mu)^*}{U}
        \Delta_\mathrm{A}
        - \frac{(v_{\mathrm{p}2, \bm A}^{\mu})^*}{U}
        \Delta_\mathrm{B}
    +  \frac{1}{N_\mathrm{s}}\frac{\partial \mathcal{E}_{0, \bm A}}{\partial A_\alpha}.
\end{align}
Evaluating this with the explicit form of $v_{\mathrm{p}1, \bm A}^{\mu}$ and $v_{\mathrm{p}2, \bm A}^{\mu}$, we obtain 
\begin{align}
    \bm j_{\mathrm{p}, \bm A}
    &= -\frac{4J_2}{U^2} \cos(2\phi) \left(|\Delta_{\mathrm{A}}|+ |\Delta_{\mathrm{B}}|^2 \right) \sum_{j=1}^3 \bm \beta_j \sin(- 2 \bm A \cdot \bm \beta_j) \nonumber \\
    & \quad -\frac{4J_2}{U^2} \sin (2\phi) \left( |\Delta_{\mathrm{A}}|^2- |\Delta_{\mathrm{B}}|^2\right)\sum_{j=1}^3 \bm \beta_j  \cos (- 2 \bm A \cdot \bm \beta_j)  \nonumber \\
    & \quad ~ + \frac{4 J_1}{U^2}|\Delta_{\mathrm{A}}| |\Delta_{\mathrm{B}}| \sum_{j=1}^3 \bm \alpha_j \sin(2\bm A\cdot \bm \alpha_j + \theta_\mathrm{AB}), 
\end{align}
where $\bm \alpha_1 = (1, \sqrt{3})$, $\bm \alpha_2 = (1, -\sqrt{3})$, $\bm \alpha_3 = (-2, 0)$, $\bm \beta_1 = (1, 0)$, $\bm \beta_2 = (-1/2, \sqrt{3}/2)$, $\bm \beta_3 = (-1/2, -\sqrt{3}/2)$, and $\theta_\mathrm{AB} = \arg\Delta_{\mathrm{A}} - \arg \Delta_{\mathrm{B}}$. 

To extract the Hall current, we set the direction of the applied electric field as $\bm A(t) = A(t) \bm e_{\theta}$ with $\bm e_{\theta} = (\cos \theta, \sin \theta)$, and consider the current component perpendicular to the electric field, $j_{\mathrm{p}, \bm A}^\perp = \bm j_{\mathrm{p}, \bm A} \cdot \bm e_{\theta+\frac{\pi}{2}}$. Then, we obtain
\begin{align}
    j_{\mathrm{p}, \bm A(t)}^\perp &= - \frac{2J_1}{\sqrt{3} U^2} |\Delta_{\mathrm{A}}| |\Delta_{\mathrm{B}}| \sin (\theta_\mathrm{AB}) \cos(3\theta) A(t)^2 \nonumber \\
    &\quad - \frac{6 J_2}{U^2} (|\Delta_{\mathrm{A}}|^2 - |\Delta_{\mathrm{B}}|^2) \sin (2\phi)  \sin(3\theta) A(t)^{2} \nonumber \\
    &\quad ~ + \mathcal{O}(E(t)^4). \label{eq:supp_current_expansion}
\end{align}
Based on our numerical calculation, $\theta_\mathrm{AB}$ is zero in equilibrium within the parameter regime we consider, and thus the first term in Eq.~\eqref{eq:supp_current_expansion} vanishes, while it can give a finite contribution to the higher-order responses when $\theta_\mathrm{AB}$ becomes finite dynamically. As in the main text, we consider the applied electric field in the $y$-direction, i.e., $\theta=\pi/2$, and the induced current in the $x$-direction. Through a straightforward calculation, we obtain   
\begin{align}
    \sigma^{x y y}(2\omega; \omega, \omega) &=  \frac{6 J_2}{U^2 \omega^2} (|\Delta_{\mathrm{A}}|^2 - |\Delta_{\mathrm{B}}|^2) \sin (2\phi).
\end{align}
This is the formula for the second-order Hall conductivity that arises from the pair hopping term proportional to $J_2$. Whereas the calculation in this section does not consider the effect of dissipation as mentioned before, various properties, such as the $1/\omega^2$-divergence and $\sin(2\phi)$-dependence, are consistent with the results with dissipation.

\onecolumngrid

\section{Green's function formulas for $\mathcal{A}^{\mu\nu\lambda}$ and $\mathcal{B}^{\mu\nu\lambda}$ }
\label{sec:supp_GF_formula_for_div_factors}

In this section, we present the explicit forms of the Green's function formulas for $\mathcal{A}^{\mu\nu\lambda}$ and $\mathcal{B}^{\mu\nu\lambda}$ in the main text. This can be derived by expanding the integrand in Eq.~\eqref{eq:formula_conductivity} with respect to $\omega_1$ and $\omega_2$ around $\omega_1=\omega_2=0$ and comparing it with Eq.~\eqref{eq:div_coeff}. The explicit forms are as follows:
    \begin{align*}
        \mathcal{A}^{\mu \nu \lambda}&=
         \sum_{\bm k }
         \int\!\frac{d\Omega}{2\pi i} f(\Omega) \Big\{ \frac{1}{2}  \mathrm{Tr} \left[ X^{\mu \nu \lambda}_{\bm k} [G^{A(0)}_{\bm k }(\Omega)  - G^{R(0)}_{\bm k }(\Omega)]\right] \\
        & \nspace{8} 
        -\frac{1}{2} \mathrm{Tr} \left[ 
         V^{\mu}_{\bm k} G_{\bm k }^{R(0)}(\Omega ) W_{\bm k}^{\nu \lambda}  G^{R(0)}_{\bm k }(\Omega)\right] + \frac{1}{2} \mathrm{Tr} \left[ V^{\mu}_{\bm k}  G^{A(0)}_{\bm k }(\Omega) W_{\bm k}^{\nu \lambda} G_{\bm k }^{A(0)}(\Omega ) \right]  \\
        & \nspace{8}  - \mathrm{Tr} \left[
        W_{\bm k}^{\mu \nu} G^{R(0)}_{\bm k }(\Omega)  V_{\bm k}^\lambda G^{R(0)}_{\bm k }(\Omega)\right] + \mathrm{Tr} \left[W_{\bm k}^{\mu \nu}  G^{A(0)}_{\bm k }(\Omega) V_{\bm k}^\lambda G^{A(0)}_{\bm k }(\Omega ) \right] \\
        & \nspace{8} -\mathrm{Tr} 
        \left[ V^{\mu}_{\bm k}  G_{\bm k }^{R(0)}(\Omega ) V_{\bm k}^\nu G_{\bm k }^{R(0)}(\Omega) V_{\bm k}^\lambda  G^{R(0)}_{\bm k }(\Omega) \right] 
        + \mathrm{Tr} 
         \left[V^{\mu}_{\bm k} G^{A(0)}_{\bm k }(\Omega)  V_{\bm k}^\nu G_{\bm k}^{A(0)}(\Omega) V_{\bm k}^\lambda  G_{\bm k }^{A(0)}(\Omega) \right] \\
        & \nspace{8} + [\nu \leftrightarrow \lambda] \Big\} + \frac{1}{2} \mathcal{E}_0^{\mu\nu\lambda},
    \end{align*}
    \vspace{-0.5cm}
    \begin{align*}
        \mathcal{B}^{\mu \nu \lambda}&=
       \sum_{\bm k }
         \int\!\frac{d\Omega}{2\pi i} f(\Omega) \left\{ 
            \mathrm{Tr} \left[ 
            V^{\mu}_{\bm k} 
            F_{\bm k }^{R(0)}(\Omega )
            W_{\bm k}^{\nu \lambda}  
            [G^{A(0)}_{\bm k }(\Omega) - G^{R(0)}_{\bm k }(\Omega)] 
            - V^{\mu}_{\bm k} [G^{A(0)}_{\bm k }(\Omega) - G^{R(0)}_{\bm k }(\Omega)] W_{\bm k}^{\nu \lambda} F_{\bm k }^{A(0)}(\Omega ) \right] \right. \\
        &\nspace{8.5}  + \mathrm{Tr} \left[
        W_{\bm k}^{\mu \lambda}  F^{R(0)}_{\bm k }(\Omega) V_{\bm k}^\nu [G^{A(0)}_{\bm k }(\Omega) - G^{R(0)}_{\bm k }(\Omega)] 
        - W_{\bm k}^{\mu \lambda}  [G^{A(0)}_{\bm k }(\Omega)  - G^{R(0)}_{\bm k }(\Omega)] V_{\bm k}^\nu F^{A(0)}_{\bm k }(\Omega) \right] \\
        & \nspace{8.5} +\mathrm{Tr} 
        \left[V^{\mu}_{\bm k} F_{\bm k}^{R(0)}(\Omega) V_{\bm k}^{\nu} G^{A(0)}_{\bm k }(\Omega) V_{\bm k}^{\lambda} G_{\bm k }^{A(0)}(\Omega )  - V^{\mu}_{\bm k} F_{\bm k }^{R(0)}(\Omega ) V_{\bm k}^\nu G_{\bm k }^{R(0)}(\Omega ) V_{\bm k}^\lambda G^{R(0)}_{\bm k }(\Omega) \right. \\
        & \nspace{11} 
        -  V^{\mu}_{\bm k} [G^{A(0)}_{\bm k }(\Omega) - G^{R(0)}_{\bm k }(\Omega)] V_{\bm k}^\nu F_{\bm k}^{A(0)}(\Omega) V_{\bm k}^\lambda  G_{\bm k }^{A(0)}(\Omega) 
        \\
        &\left. \nspace{11} -  V^{\mu}_{\bm k} [G^{A(0)}_{\bm k }(\Omega)  - G^{R(0)}_{\bm k }(\Omega)] V_{\bm k}^\nu G_{\bm k}^{A(0)}(\Omega) V_{\bm k}^\lambda F_{\bm k }^{A(0)}(\Omega) \right]  \\
        & \nspace{8.5} +\mathrm{Tr} 
        \left[ 
         V^{\mu}_{\bm k} G_{\bm k}^{R(0)}(\Omega ) V_{\bm k}^{\lambda} G^{R(0)}_{\bm k }(\Omega)  V_{\bm k}^{\nu} F_{\bm k }^{A(0)}(\Omega)  
        -  V^{\mu}_{\bm k} G^{A(0)}_{\bm k }(\Omega)  V_{\bm k}^\lambda G_{\bm k}^{A(0)}(\Omega ) V_{\bm k}^\nu  F_{\bm k }^{A(0)}(\Omega ) \right. \\
        & \nspace{11} +V^{\mu}_{\bm k}  F_{\bm k }^{R(0)}(\Omega) V_{\bm k}^\lambda G_{\bm k }^{R(0)}(\Omega) V_{\bm k}^\nu (G^{A(0)}_{\bm k }(\Omega)  - G^{R(0)}_{\bm k }(\Omega)) \\
        &
        \left. 
        \nspace{11}
        + V^{\mu}_{\bm k}  G_{\bm k }^{R(0)}(\Omega ) V_{\bm k}^\lambda F_{\bm k }^{R(0)}(\Omega) V_{\bm k}^\nu (G^{A(0)}_{\bm k }(\Omega)  - G^{R(0)}_{\bm k }(\Omega)) \right]. 
    \end{align*} 
    We have used these formulas to obtain the data shown in Fig.~\ref{fig:phase_temp_dep} in the main text. Note that the same formulas except for the energy shift term $\propto \mathcal{E}_0^{\mu\nu\lambda}$ are already shown in Ref.~\cite{Watanabe2022}. 

\vspace{0.5cm}

\twocolumngrid

\section{Floquet engineering of a pair hopping term with a geometric phase}
\label{sec:supp_Floquet}

In this section, we show that the pair-hopping term with a complex phase factor can be realized via Floquet engineering~\cite{Bukov2015, Oka2019, Weitenberg2021}. This is a straightforward generalization of the method used to implement the Haldane model~\cite{Oka2009,Jotzu2014,McIver2020}. 

We consider the following time-dependent Hamiltonian:
\begin{align}
    \hat{H}(t) &= \sum_{i,j=1}^N \sum_{\sigma=\uparrow, \downarrow} t_{i,j}(t) \hat{c}^\dagger_{i,\sigma} \hat{c}_{j,\sigma} \nonumber \\
    &\quad  + \sum_{i,j=1}^N J_{i,j}(t) \hat{c}^\dagger_{i,\uparrow} \hat{c}^\dagger_{i,\downarrow} \hat{c}_{j,\downarrow} \hat{c}_{j,\uparrow}  - U \sum_{i=1}^N \hat{n}_{i,\uparrow} \hat{n}_{i,\downarrow}, 
\end{align}
where $t_{i,j}(t)$ and $J_{i,j}(t)$ are the single-particle hopping and pair hopping amplitudes depending on time $t$, respectively. We assume that they are time periodic with a period $T$, i.e., $t_{i,j}(t+\mathcal{T}) = t_{i,j}(t)$ and $J_{i,j}(t+\mathcal{T}) = J_{i,j}(t)$. 
We consider the effective Hamiltonian and its high-frequency expansion~\cite{Bukov2015,Oka2019}. The effective Hamiltonian up to the $1/\omega$-order is given by

\begin{equation}
    \hat{H}_\mathrm{eff} = \hat{H}^{(0)} + \sum_{n=1}^{\infty} \frac{[\hat{H}^{(n)}, \hat{H}^{(-n)}]}{n\omega}, 
\end{equation}
where $\hat{H}^{(n)}$ is the $n$-th Fourier component of the Hamiltonian defined as
\begin{equation}
    \hat{H}(t) = \sum_{n=-\infty}^{\infty} \hat{H}^{(n)} e^{i n \omega t},
\end{equation}
with $\omega = 2\pi/T$. The explicit form of $\hat{H}^{(n)}$ is given by
\begin{align}
    \hat{H}^{(n)} &= \sum_{i,j=1}^N \sum_{\sigma=\uparrow, \downarrow} t_{i,j}^{(n)} \hat{c}^\dagger_{i,\sigma} \hat{c}_{j,\sigma} \nonumber \\
    &\quad  + \sum_{i,j=1}^N J_{i,j}^{(n)} \hat{c}^\dagger_{i,\uparrow} \hat{c}^\dagger_{i,\downarrow} \hat{c}_{j,\downarrow} \hat{c}_{j,\uparrow} - U \delta_{n,0} \sum_{i=1}^N \hat{n}_{i,\uparrow} \hat{n}_{i,\downarrow},
\end{align}
where $t_{i,j}^{(n)}$ and $J_{i,j}^{(n)}$ are the Fourier components of $t_{i,j}(t)$ and $J_{i,j}(t)$, respectively. Calculating the commutator in $\hat{H}_\mathrm{eff}$, we obtain the effective Hamiltonian as
\begin{align}
    \hat{H}_\mathrm{eff} &= \sum_{i,j=1}^N \sum_{\sigma=\uparrow, \downarrow} t_{i,j}^{(0)} \hat{c}^\dagger_{i,\sigma} \hat{c}_{j,\sigma} + \sum_{i,j=1}^N J_{i,j}^{(0)} \hat{c}^\dagger_{i,\uparrow} \hat{c}^\dagger_{i,\downarrow} \hat{c}_{j,\downarrow} \hat{c}_{j,\uparrow} \nonumber \\
    &\quad - U  \sum_{i=1}^N \hat{n}_{i,\uparrow} \hat{n}_{i,\downarrow} + \hat{H}_\text{LISH} + \hat{H}_\text{LIPH} + \hat{H}_{\text{3-sites}},
\end{align}
where
\begin{gather}
    \hat{H}_\text{LISH} = \sum_{i,j=1}^N \sum_{\sigma=\uparrow, \downarrow} \tilde{t}_{i,j} \hat{c}^\dagger_{i,\sigma} \hat{c}_{j,\sigma},  \\
    \hat{H}_\text{LIPH} = \sum_{i,j,k = 1}^N  \tilde{J}_{i,j, k} \hat{c}^\dagger_{i,\uparrow} \hat{c}^\dagger_{i,\downarrow} (1 - \hat{n}_{k}) \hat{c}_{j,\downarrow} \hat{c}_{j,\uparrow}, \label{eq:supp_ham_LIPH} \\
    \hat{H}_{\text{3-sites}} =\sum_{i,j,k = 1}^N  \left[ \tilde{V}_{i,j,k,i} \hat{c}^\dagger_{k \uparrow} \hat{c}^\dagger_{k \downarrow} (\hat{c}_{i,\uparrow} \hat{c}_{j,\downarrow} - \hat{c}_{i,\downarrow} \hat{c}_{j,\uparrow} ) + \mathrm{h.c.} \right],\\
    \tilde{t}_{i,j} = \sum_{n=1}^{\infty} \sum_{k=1}^N \frac{t_{i,k}^{(n)} t_{k,j}^{(-n)} - t_{i,k}^{(-n)}t_{k,j}^{(n)}}{n\omega}, \\
    \tilde{J}_{i,j, k} = \sum_{n=1}^{\infty} \sum_{k=1}^N \frac{J_{i,k}^{(n)} J_{k,j}^{(-n)} - J_{i,k}^{(-n)}J_{k,j}^{(n)}}{n\omega}, \label{eq:supp_J_tilde} \\
    \tilde{V}_{i,j,k,l} = \sum_{n=1}^{\infty} \frac{ t_{i,j}^{(n)}J_{k,l}^{(-n)}- t_{i,j}^{(-n)} J_{k,l}^{(n)}}{n\omega}.
\end{gather}
On top of the renormalization of the orginal hoppings, $t_{i,j}(t) \to t_{i,j}^{(0)}$ and $J_{i,j}(t) \to J_{i,j}^{(0)}$, there appear additional three terms:  the light-induced single-particle hopping term $\hat{H}_\text{LISH}$, the light-induced pair hopping term $\hat{H}_\text{LIPH}$, the three-sites term $\hat{H}_{\text{3-sites}}$. 

To obtain the explicit form of $\hat{H}_\text{LIPH}$, we consider the driving with a circularly polarized laser field,
\begin{equation}
    \bm A(t) = A_0 \cos(\omega t) \bm e_x + A_0 \sin(\omega t) \bm e_y,
\end{equation}
where $\bm e_x$ and $\bm e_y$ are the unit vectors in the $x$ and $y$ directions, respectively.
For this case, $J_{i,j}(t)$ is given by
\begin{equation}
    J_{ij}(t) 
    = e^{- 2 i \bm A(t) \cdot \bm r_{ij}} J_{ij}, \label{eq:supp_J_ij}
\end{equation}
where $\bm r_{ij}$ is the vector connecting the $i$-th and $j$-th sites.  Using \eqref{eq:supp_J_ij}, we obtain the Fourier components as
\begin{equation}
    J_{ij}^{(n)} = e^{- i \left(\theta_{ij}+\frac{\pi}{2}\right) n} J_n(2 A_0 r_{ij}) J_{ij}, \label{eq:supp_J_ij_n}
\end{equation}
where $\theta_{ij}$ is the angle between $\bm r_{ij}$ and the $x$-axis. 
For simplicity, we assume that there only exist nearest-neighbor couplings in $J_{i,j}(t)$. Then, using Eqs.~\eqref{eq:supp_ham_LIPH}, \eqref{eq:supp_J_tilde}, and \eqref{eq:supp_J_ij_n}, we obtain the explicit form of $\hat{H}_\text{LIPH}$ as
\begin{equation}
    \hat{H}_\text{LIPH} = \mathcal{F}(A_0)  \!\! \sum_{\llangle i, j \rrangle_{k}} e^{2i\phi} \hat{c}^\dagger_{i,\uparrow} \hat{c}^\dagger_{i,\downarrow}  (1 - \hat{n}_{k}) \hat{c}_{j,\downarrow} \hat{c}_{j,\uparrow} + \mathrm{h.c.}, \label{eq:supp_explicit_LIPH}
\end{equation}
where $\phi=\pi/4$ and
\begin{equation}
    \mathcal{F}(A_0) = \sum_{n=1}^\infty \frac{2 (J_n(2 A/\sqrt{3}))^2 }{n \omega} \sin\left(\frac{2 n \pi}{3}\right). \label{eq:supp_coeff_F} 
\end{equation}
Here, $\llangle i, j \rrangle_{k}$ denotes that the $i$-th and $j$-the sites are the next-nearest neighbor sites connected through the $k$-th site. 
Since the coefficient \eqref{eq:supp_coeff_F} is generally nonzero, Eq.~\eqref{eq:supp_explicit_LIPH} clearly shows that the pair hopping term with a complex phase factor is possible. 
While the phase factor is limited to a specific value in this case, it is possible to realize the pair hopping term with a general phase factor by considering the preexisting next-nearest neighbor coupling in $J_{ij}(t)$ additionally~\cite{Jotzu2014}.

$\hat{H}_\text{LIPH}$ contains $(1-\hat{n}_k)$, which does not appear in the pair hopping discussed in the main text. 
In the mean-field picture, one can replace it with $(1-\langle\hat{n}_k\rangle)$, which depends on the density.
Although this factor will induce quantitative differences, it is expected that it does not change the properties related to the geometric phases, since it only gives a real-valued factor. Also, this factor does not change both spatial inversion and time-reversal symmetries, which are essential to realize the superconducting nonlinear Hall effect as discussed in the main text based on the Ginzburg-Landau analysis, and thus it is expected not to affect whether the superconducting nonlinear Hall effect can be realized or not. 

\end{document}